\documentclass[aps,prb,preprint,showpacs,superscriptaddress,floatfix]{revtex4}
\usepackage{graphics}
\usepackage{epsfig}
\usepackage{amsmath}
\usepackage{amssymb}
\usepackage{calc}
\newcommand{\beq}{\begin{equation}}
\newcommand{\eeq}{\end{equation}}
\newcommand{\bea}{\begin{eqnarray}}
\newcommand{\eea}{\end{eqnarray}}

\begin{document}
\title{
STU Black Holes as Four Qubit Systems}
\author{P\'eter L\'evay}
\affiliation{Department of Theoretical Physics, Institute of Physics, Budapest 
University of Technology, H-1521 Budapest, Hungary}
\date{\today}
\begin{abstract}
In this paper we describe the structure of extremal stationary spherically symmetric black hole solutions
in the $STU$ model of $D=4$, $N=2$ supergravity
in terms of four-qubit systems.
Our analysis extends the results of previous investigations based on three
 qubits.
 The basic idea facilitating this four-qubit interpretation is the fact
 that stationary solutions in $D=4$ supergravity
 can be described by dimensional reduction along the time direction.
In this $D=3$ picture the global symmetry group $SL(2,\mathbb R)^{\times 3}$ of the model is extended by
the Ehlers $SL(2,{\mathbb R})$ accounting for the fourth qubit.
We introduce a four qubit state depending on the charges (electric, magnetic and NUT) the moduli and the warp factor.
We relate the entanglement properties of this state to different classes of black hole solutions in the STU model.
In the terminology of four qubit entanglement extremal black hole solutions
correspond to nilpotent, and nonextremal ones to semisimple states.
In arriving at this entanglement based scenario the role of the four algebraically independent four qubit $SL(2,\mathbb C)$ invariants is emphasized.

\end{abstract}
\pacs{
11.25.Mj, 03.65.Ud, 03.67.Mn, 04.70.Dy}
\maketitle{}

\section{Introduction}

Recently striking multiple relations have been discovered between two
seemingly unrelated fields: Quantum Information Theory (QIT) and the physics of black hole solutions in String Theory\cite{Duff2,Linde,Levay1}.
Although the physical basis for this 
black hole qubit correspondence (or black hole analogy) is still to be clarified, it has repeatedly proved
to be useful for obtaining additional insight into one of the two fields by exploiting methods and techniques of the other\cite{DF1,Levay2,DF2,Levay3,Scherbakov,Borsten1,stu,Borsten2,Borsten3}. 
The main correspondence found\cite{Duff2,Linde,DF1,Levay2,DF2,LVS} is between the macroscopic entropy
formulas obtained for certain black hole solutions in supergravity theories and
multiqubit and qutrit entanglement measures used in Quantum Information Theory.
The basic reason for this correspondence is the occurrence of similar groups of symmetry in these very different contexts.
On the stringy black hole side the groups in question are the global symmetry groups of $D=4$ classical supergravities, and on the QIT one the groups of local transformations for entangled subsystems {\it not} changing their multipartite entanglement. 
As far as physics is concerned an attempt has been made to understand these mathematical coincidences in terms of wrapped brane configurations giving rise to 
qubits\cite{Borsten1}.

Apart from understanding black hole entropy in quantum information theoretic terms the desire for an entanglement based understanding for issues of dynamics
also 
arose.
In particular in the special case of the STU model\cite{Duff1} it has been realized\cite{Levay1} that it is possible to rephrase the attractor mechanism\cite{attractor} as a distillation procedure of entangled "states" of very special kind on the event horizon.
Such "states" for $D=4$ extremal static spherical symmetric solutions are arising from more general ones of the form\cite{Levay1,Levay3,Levszal}
\beq
\qquad\vert\psi(\tau)\rangle\equiv ({\cal V}\otimes {\cal V}\otimes                   {\cal V})(S_3(\tau)\otimes S_2(\tau)\otimes S_1(\tau))\vert\gamma\rangle,
\qquad \tau\equiv \frac{1}{r}.
\label{3qubitallapot}
\eeq
Here
\beq
{\cal V}\equiv\frac{1}{\sqrt{2}}\begin{pmatrix}i&-1\\i&1\end{pmatrix},\qquad
S_j\equiv\frac{1}{\sqrt{y_j(\tau)}}\begin{pmatrix}y_j(\tau)&0\\-x_j(\tau)&1\end{pmatrix},\qquad j=1,2,3
\label{Smatrix}
\eeq
\beq
\vert\gamma\rangle=\sum_{a_3,a_2,a_1=0,1}{\gamma}_{a_3a_2a_1}\vert a_3a_2a_1\rangle\qquad \vert a_3a_2a_1\rangle\in{\mathbb C}^2\otimes{\mathbb C}^2\otimes {\mathbb C}^2
\eeq                                                                            \noindent
\beq                                                                            
\begin{pmatrix}{\gamma}_{000},&{\gamma}_{001},&{\gamma}_{010},&{\gamma}_{100}\\{\gamma}_{111},&{\gamma}_{110},&{\gamma}_{101},&{\gamma}_{011}\end{pmatrix}
\equiv
\frac{1}{\sqrt{2}}\begin{pmatrix}p^0,&p^1,&p^2,&p^3\\-q_0,&q_1,&q_2,&q_3
\label{gammacharge}
\end{pmatrix},                                                                  \eeq                                                                            \noindent                                                                       
where $r$ is the radial distance from the event horizon, $z_j(\tau)=x_j(\tau)-iy_j(\tau)$, $j=1,2,3$ are the scalar fields, $q_I$, and $p^I$, $I=0,1,2,3$ are the electric and magnetic charges occurring in the STU model\cite{stu}. As we see these quantities are organized into a ${\it complex}$ three-qubit state. 
This instructive notation clearly expresses the triality symmetry of the STU model\cite{Duff1}.
Moreover, the classical symmetry group of the model (i.e. $SL(2,{\mathbb R})^{\times 3}$) is manifested in this formalism by the fact that apart from the unitary matrices ${\cal V}$, $\vert\psi\rangle$ is lying on the $SL(2,{\mathbb R})^{\times 3}$ orbit of the "charge-state" $\vert\gamma\rangle$.
The unitaries ${\cal V}^{\otimes 3}$ provide an embedding of the $SL(2,{\mathbb R})^{\times 3}$ symmetry group of this $N=2$ supergravity model into $GL(2,{\mathbb C})^{\times 3}$.

The state of Eq.(\ref{3qubitallapot}) has a number of remarkable properties\cite{Levay1,Levay3,Levszal}.

$1.$ The three-tangle\cite{Kundu} $\tau_{123}$, the unique triality and $SL(2,{\mathbb C})^{\times 3}$ invariant $3$-qubit entanglement measure 
based on 
Cayley's hyperdeterminant\cite{Cayley,Zelevinsky},
for $\vert\psi\rangle$ is 
related to the macroscopic black hole entropy in the STU model as
\beq
S=\pi\sqrt{{\tau}_{123}(\vert\psi\rangle)}={\pi}\sqrt{\tau_{123}(\vert\gamma\rangle)}.
\eeq
\noindent

$2.$ The norm of $\vert\psi\rangle$ with respect to the usual scalar product in ${\mathbb C}^8$ with complex conjugation in the first factor is the Black Hole Potential\cite{stu} $V_{BH}$.

 $3.$ The flat covariant derivatives with respect to the K\"ahler connection
are acting on $\vert\psi\rangle$ as bit flip errors on the qubits. 

$4.$ 
For $BPS$-solutions and for non-BPS solutions with vanishing central charge\cite{Scherbakov} 
$\vert\psi(\infty)\rangle$ is a $GHZ$-state\cite{GHZ}.
For non-BPS solutions with non-vanishing central charge the corresponding states are graph-states
known from QIT\cite{graph}.
In this respect moduli stabilization is related to a distillation procedure of  states with special entanglement properties at the event horizon.

$5.$ On the horizon bit flip errors on $\vert\psi\rangle$ are supressed for BPS solutions
and for non-BPS ones they are not. The non-BPS solutions can be characterized
by the number and types of bit-flip errors.

$6.$ After solving the equations of motion one obtains the attractor flow $z_j(\tau)$ in moduli space.
There is a flow $\vert\psi(\tau)\rangle$ associated to this one.
For the non-BPS seed solution\cite{seed} it is possible to study how the distillation procedure unfolds itself\cite{Levszal} with the following result. In the asymptotically flat region
we are starting with a $\vert\psi(0)\rangle$ having $7$ {\it nonequal} nonvanishing amplitudes and finally at the horizon we get a graph state $\vert\psi(\infty)\rangle$
with merely $4$  nonvanishing ones with {\it equal} magnitudes. 

$7.$
The magnitude of the nonvanishing amplitudes of such "attractor states" is proportional to the black hole entropy.
The relative phases of the amplitudes reflect the structure of the fake superpotential.

$8.$ If we are starting with the very special values for the moduli corresponding to flat directions\cite{flat} this uniform structure at the horizon deteriorates\cite{Levszal}, 
with the interpretation of errors of more general types acting  on the qubits of the relevant attractor states.

In addition to these interesting results based on {\it three-qubit} states
there are ones which strongly hint at the possibility that for a complete understanding of STU black holes we have to embed our three-qubit states into {\it four-qubit ones}\cite{Levszal}.
In particular one can generalize Eq.(\ref{3qubitallapot}) by also including the warp factor $U(\tau)$
occurring in the static, spherically symmetric ansatz for the $4D$ space-time metric
\beq
ds^2=-e^{2U(\tau)}dt^2+e^{-2U(\tau)}d{\bf x}^2
\label{metricansatz}
\eeq
\noindent
into a new state $\vert\chi\rangle$ defined as\cite{Levszal}
\beq
\vert\chi(\tau)\rangle=e^{U(\tau)}\vert\psi(\tau)\rangle.
\label{chi}
\eeq
\noindent
For the non-BPS seed solution\cite{seed} it has been shown that the $7$ nonvanishing $\tau$ dependent amplitudes of this state depending on the charges, the moduli and the warp factor satisfy a system of {\it first order} differential equations.
This finding conforms with recent work done within the framework of the first order formalism for non-BPS solutions based on the so called {\it fake superpotential}\cite{fake}.

Moreover, 
within the realm of the more general class of stationary solutions
it is well-known that the warp factor taken together with the NUT potential\cite{NUT} ${\sigma}$ forms another $SL(2,{\mathbb R})$ doublet , a doublet with respect to the Ehlers group\cite{Ehlers}. 
Hence it is natural to suspect that for stationary solutions objects like $\vert\chi(\tau)\rangle$
are really four-qubit states in disguised form with the Ehlers group acting on a hidden extra qubit. The properties of these hypothetical $4$-qubit states should account for the first order formalism hiding behind the integrability of the non-BPS flow equations.

Recent investigations clearly demonstrated that this should indeed be the case\cite{Padi,Galtsov,Bossard,Stelle,Stelle2,Pioline1,Trigiante,Trigi}.
The key observation is that stationary solutions in $D=4$ supergravity
can be elegantly described by dimensional reduction along the time direction\cite{Gibbons}.
In this picture stationary solutions can be identified as solutions to a $D=3$ non-linear sigma model with target space being a symmetric space $G/H$ with $H$ {\it non-compact}.
The property that is of basic significance for us is that the group $G$ in this case extends the global symmetry group $G_4$ of $D=4$ supergravity, by also incorporating the Ehlers $SL(2,{\mathbb R})$.
In our specific case the $N=2$ STU model can be regarded as a consistent truncation of maximal $N=8$, $D=4$ supergravity with $G_4= E_{7(7)}$, truncating to $SL(2,{\mathbb R})^{\times 3}$.
Timelike reduction in the general case then yields the coset $E_{8(8)}/SO^{\ast}(16)$,
or in the case of the STU truncation the one ${\cal M}_3=SO(4,4)/SL(2,{\mathbb R})^{\times 4}$. We then expect the four copies of $SL(2,{\mathbb R})$s giving rise to the group of local operations acting on the four qubits.
Here the fourth qubit which accounts for the Ehlers group will then play a special role.

The manifold ${\cal M}_3$ is the target space of the aforementioned sigma model.
It has been proved\cite{Trigiante} that for such symmetric target spaces stationary spherically symmetric black hole solutions
can be obtained as geodesic curves on this pseudo Riemannian target space.
Such geodesic curves are classified in terms of the Noether charges of the solutions.
In the case of the STU model the coset representative ${\cal P}$ of our target space ${\cal M}_3$ and the related Noether charge can be written in the form reminiscent of a $4$-qubit state\cite{Trigiante,Bossard,Trigi}.
Moreover the line element on ${\cal M}_3$ can be written in the form\cite{Bossard} ${\rm Tr}({\cal P}^2)$ which turns out to be just the quadratic 
$4$-qubit invariant, one of the four algebraically independent invariants characterizing $4$-qubit systems\cite{Luque}.
It has also been observed\cite{Trigiante,Trigi} that the entanglement properties of such $4$-qubit-like states seem to be related to the fact whether the extremal solution in question is BPS, non-BPS or non-BPS with vanishing central charge.
Based on this finding the authors of this paper\cite{Trigiante} mention that there might be a connection with issues concerning the black hole qubit correspondence, though in this field the $D=3$ reformulation has never been used. (See Egs. (5.51)-(5.52) of that paper.)

The aim of the present paper is to show, that using the $D=3$ picture such  $4$-qubit interpretation indeed  emerges naturally.
Moreover, after establishing the desired connection we see that in this framework many aspects 
of the usual three-qubit interpretation can be understood in a nice and unified way.  

The organization of this paper is as follows.
In Section II. we present the background material on the STU model and the basics of the $D=3$ picture emerging after reduction along the time direction.
In Section III. in a four-qubit notation we reconsider the usual Iwasawa parametrization of the physical patch of the pseudo-Riemannian manifold ${\cal M}_3$.
This formalism is exploited in Section IV. where we describe the line element on ${\cal M}_3$ as the canonical quadratic $4$-qubit $SL(2,{\mathbb C})^{\times 4}$ invariant.
Here
after a sequence of $4$-qubit transformations (Hadamard gates, phase gates, and permutations)
a very convenient realization for the "vierbein" ${\cal P}$ is obtained.
These transformations correspond to a special choice of basis in  $T_{\mathbb C}{\cal M}_3$ similar to the ones used in Ref.\cite{Bossard} rendering the "quaternionic vierbein" covariantly constant with respect to the spin connection.
In Section V. we discuss the structure of conserved charges in an entanglement based framework. Here we see how our remarkable three-qubit state of Eq. (\ref{3qubitallapot})
originates from the geometric data on ${\cal M}_3$.
As an important generalization we write down a generalization of Eq.(\ref{3qubitallapot}) for stationary solutions when the NUT charge is not zero.
Section VI. is devoted to an analysis of the static, spherically symmetric solutions. Our treatment is based on the algebraically independent $4$-qubit $SL(2,{\mathbb C})^{\times 4}$ invariants.
It is shown that in the language of QIT extremal solutions correspond to {\it nilpotent}, and nonextremal ones to {\it semisimple} $4$-qubit states. 
Nilpotent states are the ones for which all of the four algebraically independent invariants vanish.
This picture is dual to the usual characterization in terms of nilpotent orbits.
Next in this entanglement based approach a study of the usual BPS and non-BPS
solutions with vanishing central charge, and the non-BPS seed solution is given.
These investigations culminate in establishing an explicit connection between the results of Ref.\cite{Trigiante} and some standard ones on four-qubit entangled systems in QIT.
Finally we present our conclusions and comments in Section VII. In
an Appendix for the convenience of the reader we also included some background material concerning
four-qubit systems.

\section{The STU model}

In the following we consider ungauged $N=2$ supergravity in $d=4$ coupled to    $n$ vector multiplets.                                                          The $n=3$ case corresponds to the $STU$ model.                                                        
The bosonic part of the action (without hypermultiplets) is

\begin{eqnarray}
\label{action11}
{\cal S}&=&\frac{1}{16\pi}\int d^4x\sqrt{\vert g\vert }\{-\frac{R}{2}+G_{i\overline{j}}{\partial}_{\mu}z^i{\partial}_{\nu}{\overline{z}}^{\overline{j}}g^{\mu\nu}\nonumber\\&+&({\rm Im}{\cal N}_{IJ}{\cal F}^I\cdot{\cal F}^J+{\rm Re}{\cal N}_{IJ}{\cal F}^I\cdot{^\ast{\cal F}^J})\}
\end{eqnarray}
\noindent
Here ${\cal F}^I$, and ${^\ast{\cal F}^I}$,  $I=0,1,2\dots n$ are two-forms associated to the field strengths ${\cal F}^I_{\mu\nu}$  of $n+1$ $U(1)$ gauge-fields and their duals.

The $z^i$ $i=1,2\dots n$ are complex scalar (moduli) fields that can be regarded as local coordinates on a projective special K\"ahler manifold. This manifold for the STU model is
$[SL(2, \mathbb R)/U(1)]^{\times 3}$.
In the following we will denote the three complex scalar fields as
\beq
z^j\equiv x^j-iy^j,\qquad j=1,2,3,\qquad y^j>0.
\label{moduli}
\eeq
\noindent
With these definitions the metric and the connection on the scalar manifold are
\beq
G_{i\overline{j}}=\frac{\delta_{i\overline{j}}}{(2y^i)^2},\qquad
\qquad{\Gamma}^{j}_{jj}=\frac{-i}{y^j}.
\label{targetmetric}
\eeq
\noindent
The metric above can be derived from the K\"ahler potential
\beq
K= -\log(8y_1y_2y_3)
\label{Kahler}
\eeq
\noindent
as $G_{i\overline{j}}={\partial}_i{\partial}_{\overline{j}}K$.
For the STU model the scalar dependent vector couplings ${\rm Re}{\cal N}_{IJ}$
and ${\rm Im}{\cal N}_{IJ}$ take the following form
\beq
\nu_{IJ}\equiv{\rm Re}{\cal N}_{IJ}=\begin{pmatrix}2x_1x_2x_3&-x_2x_3&-x_1x_3&-x_1x_2\\-x_2x_3&0&x_3&x_2\\-x_1x_3&x_3&0&x_1\\-x_1x_2&x_2&x_1&0\end{pmatrix},
\label{valos}
\eeq
\noindent
\beq
{\mu}_{IJ}\equiv{\rm Im}{\cal N}_{IJ}=-y_1y_2y_3\begin{pmatrix}1+{\left(\frac{x_1}{y_1}\right)}^
2+{\left(\frac{x_2}{y_2}\right)}^2+{\left(\frac{x_3}{y_3}\right)}^2&-\frac{x_1}{
y_1^2}&-\frac{x_2}{y_2^2}&-\frac{x_3}{y_3^2}\\-\frac{x_1}{y_1^2}&\frac{1}{y_1^2}
&0&0\\-\frac{x_2}{y_2^2}&0&\frac{1}{y_2^2}&0\\-\frac{x_3}{y_3^2}&0&0&\frac{1}{y_
3^2}\end{pmatrix},
\label{kepzetes}
\eeq
\noindent
\beq
\mu^{IJ}\equiv({\mu}^{-1})_{IJ}=\frac{-1}{y_1y_2y_3}\begin{pmatrix}1&x_1&x_2&x_3\\x_1&{\vert z_1\vert}^2&x_1x_2&x_1x_3\\x_2&x_1x_2&{\vert z_2\vert}^2&x_2x_3\\x_3&x_1x_3&x_2x_3&{\vert z_3\vert}^2\end{pmatrix}.
\label{muinv}
\eeq
We note that these vector couplings can be derived from the
holomorphic prepotential
\beq
F(X)=\frac{X^1X^2X^3}{X^0},\qquad X^I=(X^0,X^0z^a),
\eeq
\noindent
via the standard procedure characterizing special K\"ahler geometry\cite{Dauria}.

Our aim is to describe stationary solutions of the Euler-Lagrange equations arising from the Lagrangian of the STU model in a four-qubit entanglement based language. It is well-known that the most general ansatz for stationary solutions in four dimensions is
\beq
ds^2=-e^{2U}(dt+\omega)^2+e^{-2U}h_{ab}dx^adx^b,
\label{ans1}
\eeq
\noindent
\beq
{\cal F}^{I}=d{\cal A}^{I}=d(\xi^I(dt+\omega)+ A^I),
\label{ans2}
\eeq
\noindent
where $a,b=1,2,3$ correspond to the spacial directions. The quantities $U$,     $\xi^I$, $A^I_a$, $\omega_a$ and $h_{ab}$ are regarded as $3D$ fields, i.e.
the ansatz above corresponds to dimensional reduction to $D=3$ along the timelike direction.
In achieving this we have chosen the gauge such that the Lie-derivative of ${\cal A}^I$ with respect to the timelike Killing vector vanishes, and have chosen coordinates such that the isometry corresponding to this Killing vector is just a (time) translation. In this case the quantities in Eqs.(\ref{ans1}-\ref{ans2})
are merely depending on $x^a$, $a=1,2,3$.
The ansatz for the gauge fields ${\cal A}^I$ reflects its decomposition to terms parallel ($\xi^I$), and orthogonal ($A^I$) components with respect to the timelike Killing vector\cite{Gibbons}.

After performing the dimensional reduction to $D=3$ our starting Lagrangian of
Eq. (\ref{action11}) takes the following form\cite{Gibbons,Padi}
\beq
{\cal L}={\cal L}_1+{\cal L}_2+{\cal L}_3,
\eeq
\noindent
where
\beq{\cal L}_1=-\frac{1}{2}\sqrt{h}R[h]+dU\wedge\ast dU+\frac{1}{4}e^{-4U}(d\sigma+\tilde{\xi}_Id \xi^I
-\xi^Id\tilde{\xi}_I
)\wedge\ast(d\sigma+\tilde{\xi}_Jd \xi^J
-\xi^Jd \tilde{\xi}_J
)                                  ,
\label{L1}
\eeq
\noindent
\beq
{\cal L}_2=G_{i\overline{j}}dz^i\wedge\ast d\overline{z}^{\overline{j}},
\label{L2}
\eeq
\noindent
\beq   
{\cal L}_3=\frac{1}{2}e^{-2U}\mu_{IJ}d\xi^I\wedge\ast d\xi^J+\frac{1}{2}e^{-2U}\mu^{IJ}(d\tilde{\xi}_I-\nu_{IK}d\xi^K)\wedge\ast(d\tilde{\xi}_J-\nu_{JL}d\xi^L).
\label{L3}
\eeq
\noindent
\noindent									
Here the new (axionic) scalars $\sigma$ and $\tilde{\xi}_I$ are coming from dualizing $\omega$ and $A^I$ by\cite{Gibbons} 
\beq
d\tilde{\xi}_I\equiv
\nu_{IJ}d\xi^J
-e^{2U}\mu_{IJ}\ast(dA^J+\xi^Jd\omega)
\eeq
\noindent
\beq
d\sigma\equiv e^{4U}\ast d\omega
+\xi^I d\tilde{\xi}_I
-\tilde{\xi}_Id\xi^I
.
\eeq
\noindent
Note also that here  the exterior derivative is understood on the (generally curved) spatial slice with local coordinates $x^j$, $j=1,2,3$.

The dimensionally reduced Lagrangian ${\cal L}$ can be written in the nice form
of $3D$  gravity coupled to a nonlinear sigma model defined on the spatial slice with target manifold\cite{Bossard} ${\cal M}_3=SO(4,4)/SL(2,{\mathbb R})^{\times 4}$ with the Lagrangian
\beq
{\cal L}=-\frac{1}{2}\sqrt{h}R[h]+g_{mn}{\partial}_a{\Phi}^m{\partial}^a{\Phi}^n\label{teljeslag}
\eeq
\noindent
where ${\Phi}^m, m=1,2,\dots 16$ refers to the scalar fields: $U,\sigma,\xi^I,\tilde{\xi}_I, z^j, \overline{z}^{\overline{j}}$ with $I=0,1,2,3$ and $j=1,2,3$.
Here the line element on ${\cal M}_3$ defines $g_{mn}$ as
$ds^2_{{\cal M}_3}= g_{mn}{\Phi}^m{\Phi}^n$ with the explicit form
\begin{eqnarray}
\frac{1}{4}ds^2_{{\cal M}_3}&=& G_{i\overline{j}}(z,\overline{z})dz^id\overline{z}^{\overline{j}}+dU^2+
\frac{1}{4}e^{-4U}(d\sigma+\tilde{\xi}_Id \xi^I
-\xi^Id\tilde{\xi}_I
)^2\nonumber\\&+&
\frac{1}{2}e^{-2U}\left[\mu_{IJ}d\xi^Id\xi^J+
\mu^{IJ}(d\tilde{\xi}_I-\nu_{IK}d\xi^K)(d\tilde{\xi}_J-\nu_{J
L}d\xi^L)\right].
\label{modulimetrika}
\end{eqnarray}
\noindent

In this paper we are only discussing the special case of stationary, weakly extremal solutions i.e. solutions when the spacial slices are flat\cite{Bossard,Padi}. Single centered black holes with spherical symmetry are of this type. In this case the dynamics of the moduli ${\Phi}^m$ are decoupled from the $3D$ gravity and the metric ansatz can be chosen to be the form
\beq
ds^2=-e^{2U}(dt+\omega)+e^{-2U}(dr^2+r^2(d\theta^2+{\sin}^2\theta d{\varphi}),
\eeq
\noindent
with the warp factor depending merely on $r$.
Now the equations of motion are equivalent to light-like geodesic motion on ${\cal M}_3$ with the affine parameter $\tau=\frac{1}{r}$.
Since ${\cal M}_3$ is a symmetric space there is a number of conserved Noether charges associated with this geodesic motion.
The most important ones are the electric and magnetic charges $p^I$ and $q_I$ and the $NUT$ charge $k$\cite{Bossard,Stelle,Trigiante}.
Static solutions are characterized by the vanishing of the NUT charge i.e. $k=0$. In this case the dynamics is described by the Lagrangian  of a fiducial particle in a "black-hole potential" $V_{BH}$ 

\beq
{\cal L}(U(\tau), z^i(\tau),\overline{z}^{\overline{i}}(\tau))=\left(\frac{dU}  {d\tau}\right)^2+G_{i\overline{j}}\frac{dz^i}{d\tau}\frac{d\overline{z}^{\overline{j}}}{d\tau}+e^{2U}V_{BH}(z,\overline{z},p,q),
\label{Lagrange}
\eeq
\noindent
with the constraint
\beq
\left(
\frac{d U}{d\tau}\right)^2+G_{i\overline{j}}\frac{dz^i}{d\tau}
\frac{d\overline{z}^{\overline{j}}}{d\tau}-e^{2U}V_{BH}(z,\overline{z},p,
q)=0.
\label{constraint}
\eeq

Here the black hole potential $V_{BH}$
is depending on the moduli as well on the charges.
Its explicit form is given by
\beq
V_{BH}=\frac{1}{2}\begin{pmatrix}p^I&q_I\end{pmatrix}\begin{pmatrix}(\mu+\nu{\mu}^{-1}\nu)_{IJ}&-(\nu{\mu}^{-1})^J_I\\-({\mu}^{-1}\nu)^I_J&({\mu}^{-1})^{IJ}\end{pmatrix}\begin{pmatrix}p^J\\q_J\end{pmatrix}.
\label{potential}
\eeq
\noindent
An alternative expression for $V_{BH}$ can be given in terms of the central charge of $N=2$ supergravity, i.e. the charge of the graviphoton.
\beq
V_{BH}=Z\overline{Z}+G^{i\overline{j}}(D_iZ)({\overline{D}}_{\overline{j}}\overline{Z})
\eeq
\noindent
where for the STU model

\beq
Z=e^{K/2}W=e^{K/2}(q_0+z_1q_1+z_2q_2+z_3q_3+z_1z_2z_3p^0-z_2z_3p^1-z_1z_3p^2-z_1z_2p^3),
\label{central}
\eeq
and $D_a$ is the K\"ahler covariant derivative

\beq
D_iZ=({\partial}_i+\frac{1}{2}{\partial}_iK)Z,
\label{kovika}
\eeq
\noindent
and $W$ is the superpotential.

Extremization of the effective Lagrangian Eq.(\ref{Lagrange}) with respect to the warp factor and the scalar fields yields the Euler-Lagrange equations

\beq
\ddot{U}=e^{2U}V_{BH},\qquad \ddot{z}^{i}+\Gamma^i_{jk}\dot{z}^j\dot{z}^k=e^{2U}{\partial}^iV_{BH}.
\label{Euler}
\eeq
\noindent
In these equations the dots denote derivatives with respect to $\tau=\frac{1}{r}$.
These radial evolution equations taken together with the constraint Eq.(\ref{constraint})
determine the structure of static, spherically symmetric, extremal black hole solutions in the STU model.
For the more general stationary case with nonvanishing NUT charge the motion along $\xi^I$, $\tilde{\xi}_I$ and $\sigma$ does not separate from the one on $U$ and $z^j$.
In this case we obtain the generalization of Eqs.(\ref{Euler}).
Since for our four-qubit picture we will not consider solutions of such kind we will not give the corresponding equations here.

As we have seen from this section the radial evolution associated to stationary spherical symmetric black hole solutions of the $D=4$ STU model can be described\cite{Trigiante,Trigi} as geodesic motion in the moduli space ${\cal M}_3$ 
of a dimensionally reduced $D=3$ theory.
The key issue of this reduction relevant to this paper is the enlargement of the $D=4$ symmetry group
from $SL(2,{\mathbb R})^{\times 3}$ to the $D=3$ one $SO(4,4)$ containing $SL(2,{\mathbb R})^{\times 4}$ as a subgroup.
This result paves the way for the possibility  to reinterpret our STU black holes
as four-qubit systems.

\section{The Iwasawa parametrization and four qubits}

Our starting point is the Iwasawa parametrization of the coset
${\cal M}_3=SO(4,4)/SO(2,2)\times SO(2,2)\simeq SO(4,4)/SL(2,{\mathbb R}))^{\otimes 4}$
as used in the paper of Bossard et.al.\cite{Bossard} 
For this parametrization the $16$ dimensional coset is (locally) coordinatized
by the fields $x_j, y_j$, $\phi\equiv 2U,\sigma$, and the potentials $\xi^I$ and $\tilde{\xi}_I$ quantities featuring the Lagrangian ${\cal L}$ of Eq.(\ref{teljeslag}). 
In order to avoid using disturbing factors\cite{Trigiante} of $\sqrt{2}$ 
we rescale the potentials and  define new quantities $\zeta^I$, $\tilde{\zeta}_I$ as
\beq
\zeta^I\equiv\sqrt{2}\xi^I,\qquad \tilde{\zeta}_I=\sqrt{2}\tilde{\xi}_I.
\label{renormpot}
\eeq
\noindent
In terms of these quantities the coset representative is
\beq
V\equiv e^{-\frac{1}{2}\phi H_0}\left(\prod_{j=1}^3 e^{-\frac{1}{2}\log{y_j}H_j}
e^{-x_jE_j}
\right)e^{-{\zeta}^IE_{q_I}-\tilde{\zeta}_IE_{p^I}}e^{-\sigma E_0}.
\label{Iwasawa}
\eeq
\noindent
Here the four copies of $SL(2,{\mathbb R})$ generators $H_{\alpha},E_{\alpha},F_{\alpha}$, $\alpha=0,1,2,3$
satisfy the commutation relations
\beq
[E_{\alpha},F_{\alpha}]=H_{\alpha},\quad [H_{\alpha},E_{\alpha}]=2E_{\alpha},\quad [H_{\alpha},F_{\alpha}]=-2F_{\alpha},
\label{usual}
\eeq
\noindent
and the $16$ generators of $so(4,4)$ {\it not} belonging to the $sl(2)\oplus sl(2)\oplus sl(2)\oplus sl(2)$ algebra are denoted by the symbols
$E_{p^I}, E_{q_I}, F_{p^I}, F_{q_I}$, $I=0,1,2,3$.
This decomposition of generators answers the split
\beq
so(4,4)=[sl(2,{\mathbb R})]^4\oplus (2,2,2,2)={\it h}\oplus{\it m},
\eeq
\noindent
which we would like to explicitly describe.
(For an explicit connection between our conventions described below, and the one as given by Bossard et.al.\cite{Bossard} we refer the reader to the Appendix.)

The Lie-algebra $so(4,4)$ adapted to our $4$-qubit description will be regarded
as the set of $8\times 8$ matrices ${\cal D}$ satisfying 

\beq
{\cal D}G+G{\cal D}^T=0
\label{defi}
\eeq
\noindent
where
\beq
G=\begin{pmatrix}g&0\\0&g\end{pmatrix},\quad g=\varepsilon\otimes\varepsilon\quad,\quad \varepsilon=\begin{pmatrix}0&1\\-1&0\end{pmatrix}.
\label{GG}
\eeq
\noindent
An element of $so(4,4)$ will be parametrized as
\beq
{\cal D}(s_3,s_2,s_1,s_0;D)=\begin{pmatrix}s_3\otimes I_2+I_3
\otimes s_2&Dg\\-D^Tg&s_1\otimes I_0
+I_1\otimes s_0\end{pmatrix}.
\label{fontos}
\eeq
\noindent
Here the ${\it m}$-type generators are labelled by a {\it real} $4\times 4$ matrix
\beq
D=
\begin {pmatrix}D_{0000}&D_{0001}&D_{0010}&D_{0011}\\
D_{0100}&D_{0101}&D_{0110}&D_{0111}\\
D_{1000}&D_{1001}&D_{1010}&D_{1011}\\
D_{1100}&D_{1101}&D_{1110}&D_{1111}\end {pmatrix},
\label{ime}
\eeq
\noindent
which is expressed in terms of the amplitudes of a $4$-qubit state
with index structure
\beq
D_{i_3i_2i_1i_0},\qquad i_3,i_2,i_1,i_0=0,1.
\eeq
\noindent
Notice that for convenience we have labelled the qubits from the right to the left. Moreover, the first qubit will be regarded as special explaining the somewhat unusual label: $i_0$.

The ${\it h}$ type generators are featuring the $2\times 2$  matrices $s_{\alpha}$ of the form

\beq
s_{\alpha}\equiv\begin{pmatrix}h_{\alpha}&e_{\alpha}\\f_{\alpha}&-h_{\alpha}\end{pmatrix},\qquad \alpha=0,1,2,3.
\eeq
\noindent
These matrices are expanded in terms of the ones
\beq
H=\begin{pmatrix}1&0\\0&-1\end{pmatrix},\quad
E=\begin{pmatrix}0&1\\0&0\end{pmatrix},\quad
F=\begin{pmatrix}0&0\\1&0\end{pmatrix},
\eeq
\noindent
satisfying the relations of Eq.(\ref{usual}).

The labels of the $2\times 2$ matrices appearing
in Eq. (\ref{fontos}) are referring to the qubits they act on. This action is induced by commutators of the form $[{\it h},{\it m}]\subset {\it m}$.
More precisely after commuting the block off-diagonal ${\it m}$ part with the block-diagonal ${\it h}$ one using 
\beq
s\varepsilon +\varepsilon s^T=0, \qquad s\in sl(2)
\eeq
\noindent
we get the action 
\beq
(s_3\otimes I_2+I_3\otimes s_2)D+ D(s_1^T\otimes I_0+I_1\otimes s_0^T),
\eeq
\noindent
which is the first order term in the $SL(2, {\mathbb R})^{\times 4}$ group action
\beq
D\mapsto (S_3\otimes S_2)D(S_1\otimes S_0)^T,\qquad S_{\alpha}\in SL(2, {\mathbb R}),\quad \alpha=0,1,2,3.
\label{egy}
\eeq
\noindent
Clearly this action in $4$-qubit notation reads as
\beq
D_{i_3i_2i_1i_0}\mapsto \sum_{i^{\prime}_3
i^{\prime}_2i^{\prime}_1i^{\prime}_0=0,1}
(S_3)_{i_3i^{\prime}_3}
(S_2)_{i_2i^{\prime}_2}
(S_1)_{i_1i^{\prime}_1}
(S_0)_{i_0i^{\prime}_0}
D_{i^{\prime}_3i^{\prime}_2i^{\prime}_1i^{\prime}_0},
\label{ketto}
\eeq
\noindent
or in the notation used in Quantum Information Theory
\beq
\vert D\rangle\mapsto (S_3\otimes S_2\otimes S_1\otimes S_0)\vert D\rangle
,\qquad \vert D\rangle =\sum_{i_3i_2i_1i_0=0,1}D_{i_3i_2i_1i_0}\vert i_3i_2i_1i_0\rangle.
\label{harom}
\eeq
\noindent
We remark that for the convenience of the reader in the Appendix we included 
more details on the correpondence between the structure of the group $SO(4,4)$
and $4$-qubit entanglement. 

Now returning to our coset representative  of Eq.(\ref{Iwasawa}), we introduce the new coordinates
\beq
x_0\equiv \sigma,\qquad y_0\equiv e^{\phi}=e^{2U}.
\label{ujkoord}
\eeq
\noindent
Using our $4$-qubit realization in these coordinates we have
\beq
\prod_{\alpha=0}^3e^{-\frac{1}{2}\log y_{\alpha}H_{\alpha}}e^{-x_{\alpha}E_{\alpha}}=\begin{pmatrix}M_3\otimes M_2&0\\0&M_1\otimes M_0\end{pmatrix},
\label{hpart}
\eeq
\noindent
where
\beq
M_{\alpha}\equiv\frac{1}{\sqrt{y_{\alpha}}}\begin{pmatrix}1&-x_{\alpha}\\0&y_{\alpha}\end{pmatrix}.
\label{mmatrix}
\eeq
\noindent
As a next step we introduce the $4\times 4$ matrix and its associated $4$-qubit state 
\beq
\zeta\equiv
\begin{pmatrix}{\zeta}_{0000}&{\zeta}_{0001}&{\zeta}_{0010}&{\zeta}_{0011}\\
{\zeta}_{0100}&{\zeta}_{0101}&{\zeta}_{0110}&{\zeta}_{0111}\\
{\zeta}_{1000}&{\zeta}_{1001}&{\zeta}_{1010}&{\zeta}_{1011}\\
{\zeta}_{1100}&{\zeta}_{1101}&{\zeta}_{1110}&{\zeta}_{1111}\end{pmatrix}=
\begin{pmatrix}-\tilde{\zeta}_0&0&\tilde{\zeta}_1&0\\
                            \tilde{\zeta}_2&0&{\zeta}^3&0\\
                            \tilde{\zeta}_3&0&{\zeta}^2&0\\
                            {\zeta}^1&0&{\zeta}^0&0\end{pmatrix}
\label{zetaparam}
\eeq
\noindent    
Using this we write
\beq
{\zeta}^IE_{q_I}+\tilde{\zeta}_IE_{p^I}=\begin{pmatrix}0&\zeta g\\-\zeta^Tg&0\end{pmatrix}.
\label{vissza}
\eeq
\noindent
Using the special form of the matrix $\zeta$ we have the property $\zeta g\zeta^Tg=0$ hence a staightforward calculation shows that
\beq
e^{-{\zeta}^IE_{q_I}-\tilde{\zeta}_IE_{p^I}}=\begin{pmatrix}{\bf 1}&-\zeta g\\\zeta^Tg&{\bf 1}+\frac{1}{2}\Delta\end{pmatrix},
\label{1f}
\eeq
\noindent
where ${\bf 1}\equiv I\otimes I$ and
\beq
\Delta=-\zeta^Tg\zeta g=\begin{pmatrix}{\zeta}^{(0)}\cdot{\zeta}^{(0)}&
{\zeta}^{(0)}\cdot{\zeta}^{(1)}\\{\zeta}^{(0)}\cdot{\zeta}^{(1)}&{\zeta}^{(1)}\cdot{\zeta}^{(1)}\end{pmatrix}\varepsilon\otimes E.
\label{cayleys}
\eeq
\noindent
Here the $4$-component vectors $\zeta^{(0)}$ and $\zeta^{(1)}$ are just the first and third columns of the matrix $\zeta$ of Eq.(\ref{zetaparam}),
and the $\cdot$ product is defined by Eq. (\ref{explicitalak}) of the Appendix.

Due to the special structure of $\zeta$ we also have the property
\beq
e^{x_0E_0}
e^{-{\zeta}^IE_{q_I}-\tilde{\zeta}_IE_{p^I}}e^{-x_0E_0}=
e^{-{\zeta}^IE_{q_I}-\tilde{\zeta}_IE_{p^I}},
\eeq
\noindent
resulting in our final form for the coset representative in the Iwasawa gauge
\beq
V=
\begin{pmatrix}M_3\otimes M_2&0\\0&M_1\otimes M_0\end{pmatrix}
\begin{pmatrix}{\bf 1}&-\zeta g\\\zeta^Tg&{\bf 1}+\frac{1}{2}\Delta\end{pmatrix}.                                
\label{coset}
\eeq
\noindent

We close this section with some important comments.
From the particular form of our coset representative in the Iwasawa gauge, also  reflected in our choice of the matrix $\zeta$ of Eq.(\ref{zetaparam}), we see that the role of the first qubit labelled by $i_0$ is special. The corresponding $SL(2,\mathbb R)$ action refers to the Ehlers-group.
However, our choice of $\zeta$ also gives special status to the {\it second} qubit labelled by $i_1$. This is also reflected in the structure of the matrix $\Delta$ of Eq.(\ref{cayleys}).
The $8$ components of $\zeta$ can be regarded as the ones arising from an embedding of a three-qubit state sitting inside a four-qubit one having merely $8$ nonvanishing amplitudes. The grouping of these amlitudes of this three-qubit state into two four-vectors $\zeta^{(0)}$ and $\zeta^{(1)}$
is based on the special role we have also attached to the second qubit.
However, we would have chosen any of the remaining two qubits to play this role. This would have resulted in another $4$ plus $4$ split for the $8$ nonzero components of $\zeta$.  
This freedom for different arrangements is related to the triality of $so(4,4)$ connected to the permutation symmetry inherent in the embedded three-qubit system. For more details on this point we refer the reader to the Appendix.

\section{The line element on ${\cal M}_3$ as a four-qubit invariant.}

The line element on ${\cal M}_3$ is given by the formula\cite{Bossard}

\beq
ds^2={\rm Tr}({\cal P})^2
\label{line}
\eeq
\noindent
where
\beq
{\cal P}\equiv\frac{1}{2}(dVV^{-1}+\eta (dVV^{-1})^T{\eta})
\eeq
\noindent
and the involution compatible with our conventions is
\beq
\eta=\begin{pmatrix}I\otimes I&0\\0&-I\otimes I\end{pmatrix}.
\eeq
\noindent
Using the explicit form for $V$ as given by Eq.(\ref{coset})
a straightforward calculation gives the result for ${\cal P}$
\beq
{\cal P}=\frac{1}{2}\begin{pmatrix}\Sigma_3\otimes I_2+I_3\otimes \Sigma_2&-g\Psi-\Psi g\\
g\Psi^T+\Psi^Tg&\Sigma_1\otimes I_0+I_1\otimes \Sigma_0\end{pmatrix}
\label{explicit}
\eeq
\noindent
where
\beq
\Sigma_{j}=\frac{1}{y_{j}}\begin{pmatrix}-dy_{j}&-dx_{j}\\
-dx_{j}&dy_{j}\end{pmatrix},\qquad j=1,2,3
\eeq
\noindent
and
\beq
\Sigma_{0}=\frac{1}{y_{0}}\begin{pmatrix}-dy_{0}&-dx_{0}+w\\
-dx_{0}+w&dy_{0}\end{pmatrix},\qquad w=\frac{1}{2}({\zeta}^Id\tilde{\zeta}_I-\tilde{\zeta}_Id\zeta^I).
\eeq
\noindent
The important part we have not discussed yet is the $4\times 4$ matrix
\beq
\Psi\equiv(M_3\otimes M_2)d\zeta(M_1\otimes M_0)^T,
\eeq
\noindent
which by virtue of Eqs.(\ref{egy}-\ref{harom}) can be written as a differential form on the symplectic torus determined by the Wilson lines based on a {\it four-qubit state}
\beq
\vert \Psi\rangle=(M_3\otimes M_2\otimes M_1\otimes M_0)\vert d\zeta\rangle.
\label{state}
\eeq
\noindent
Recalling our conventions of Eqs.(\ref{ujkoord}), (\ref{mmatrix}), (\ref{zetaparam})
we expect that
 $\vert \Psi\rangle$ is depending on the {\it warp factor, the NUT potential, the moduli, and the Wilson lines $d\zeta^I$ and $d\tilde{\zeta}_I$}.
The four-qubit state $\vert d\zeta\rangle$ depending only on the Wilson lines clearly
determines the entanglement type, since $\vert \Psi\rangle$ is lying on the $SL(2,{\bf \mathbb R})^{\times 4}$ orbit of this state. 
However, due to the special role of our first qubit $\vert \Psi\rangle$
is of special kind. Like in Eq.(\ref{zetaparam}) its nonzero amplitudes when displayed in a $4\times 4$ array are located in the first and the third columns.
An important consequence of this is that the NUT potential is not appearing in the explicit form of $\vert \Psi\rangle$.

We can get a four-qubit state $\vert\Phi\rangle$ of a more general type after reinterpreting the term
$g\Psi g+\Psi=(g\Psi+\Psi g)g$   
found in the upper right block of Eq.(\ref{explicit})
as a superposition
\beq
\vert \Phi\rangle= (\varepsilon\otimes\varepsilon\otimes\varepsilon\otimes \varepsilon)\vert \Psi\rangle +\vert \Psi\rangle.
\label{superposition}
\eeq
The explicit form of this state
is
\beq
\vert\Phi\rangle =(M_3\otimes M_2\otimes M_1\otimes M_0)\vert d\zeta\rangle
+
(M_3\otimes M_2\otimes M_1\otimes M_0)^{T-1}\vert d\tilde{\zeta}\rangle
\label{teljes}
\eeq
\noindent
where
$\vert\tilde{\zeta}\rangle =(\varepsilon\otimes\varepsilon\otimes\varepsilon\otimes\varepsilon)\vert\zeta\rangle$.
We see that $\vert\tilde{\zeta}\rangle$ is transforming with respect to the contragredient action. Using Eq.(\ref{zetaparam}) the explicit form of the  transformation $\vert\zeta\rangle \mapsto\vert\tilde{\zeta}\rangle$ 
is
\beq
\begin{pmatrix}-\tilde{\zeta}_0&0&\tilde{\zeta}_1&0\\
                            \tilde{\zeta}_2&0&{\zeta}^3&0\\
                            \tilde{\zeta}_3&0&{\zeta}^2&0\\
                            {\zeta}^1&0&{\zeta}^0&0\end{pmatrix}\mapsto
\begin{pmatrix}0&-{\zeta}^0&0&{\zeta}^1\\
                            0&{\zeta}^2&0&-\tilde{\zeta}_3\\
                            0&{\zeta}^3&0&-\tilde{\zeta}_2\\
                            0&-\tilde{\zeta}_1&0&-\tilde{\zeta}_0\end{pmatrix}.
\eeq
\noindent
i.e. fields with a tilde are transformed into the corresponding ones without a tilde up to some crucial signs
($\tilde{\zeta}_I\mapsto \zeta^I$ and $\zeta^I\mapsto -\tilde{\zeta}_I$), and their locations are shifted by one column.

Using these results the final form of ${\cal P}$ is
\beq
{\cal P}=\frac{1}{2}\begin{pmatrix}\Sigma_3\otimes I_2+I_3\otimes \Sigma_
2&-\Phi g\\
\Phi^Tg&\Sigma_1\otimes I_0+I_1\otimes \Sigma_0\end{pmatrix}.
\label{explicit2}
\eeq
\noindent
Using   Eqs.(\ref{line}-\ref{explicit}) we obtain for the line element the following form

\beq
ds_{{\cal M}_3}^2=\sum_{j=1}^3\frac{dx_j^2+dy_j^2}{y_j^2}+\frac{(dx_0-w)^2+dy_0^2}{y_0^2}-\vert\vert \Psi\vert\vert^2,
\label{line1}
\eeq
\noindent
where
\beq
\vert\vert \Psi\vert\vert^2\equiv\langle \Psi\vert \Psi\rangle,\qquad
\vert \Psi\rangle =(M_3\otimes M_2\otimes M_1\otimes M_0)\vert d\zeta\rangle.
\label{norma}
\eeq
\noindent

Alternatively we can consider Eq.(\ref{explicit2}) featuring $\Phi=g \Psi g+\Psi$ which is the $4\times 4$ version
of the state $\vert \Phi\rangle$. Then by virtue of the special structure of   the
matrix $\Psi$ (which is similar to the one of Eq.(\ref{zetaparam}))          satisfying
$\Psi g\Psi^Tg=0$ one gets 
\beq
\langle \Psi\vert \Psi\rangle={\rm Tr}(\Psi^T \Psi)=\frac{1}{2}{\rm Tr}(\Phi g\Phi^Tg)=\frac{1}{2}{\varepsilon}^{i_3i_3^{\prime}}
{\varepsilon}^{i_2i_2^{\prime}}{\varepsilon}^{i_1i_1^{\prime}}{\varepsilon}^{i_0i_0^{\prime}}{\Phi}_{i_3i_2i_1i_0}{\Phi}_{i_3^{\prime}i_2^{\prime}i_1^{\prime}i_0^{\prime}}
.
\label{calc}
\eeq
\noindent

We see that the term $\vert\vert \Psi\vert\vert^2$ occurring in the expression of the line element has the immediate interpretation as the norm of a four-qubit state. However, again due to the special structure of $\vert d\zeta\rangle$
which determines the orbit type of $\vert \Psi\rangle$
it is natural to give a three-qubit reinterpretation as follows.
Define
\beq
\vert \psi\rangle\equiv (M_3\otimes M_2\otimes M_1\otimes I)\vert d\zeta \rangle,\qquad
\vert \Psi\rangle =(I\otimes I\otimes I\otimes M_0)\vert \psi\rangle.
\label{ujra}
\eeq
\noindent
Then we have
\beq
\vert\vert \Psi\vert\vert^2=\frac{1}{y_0}\vert\vert \psi\vert\vert^2=e^{-2U}\vert\vert \psi\vert\vert^2,
\label{warpnorm}
\eeq
\noindent
where by virtue of
\beq
\vert d\zeta\rangle =\sum_{i_3i_2i_1=0,1}d\zeta_{i_3i_2i_10}\vert i_3i_2i_10\rangle
\eeq
\noindent
$\vert\vert \psi\vert\vert^2$ can be regarded as the norm squared of a {\it three-qubit state}.
Let us now recall Eqs.(\ref{valos})-(\ref{kepzetes}) and (\ref{muinv}).
One can check that
\beq
-e^{-2U}\begin{pmatrix}d{\zeta}^I&d\tilde{\zeta}_I\end{pmatrix}\begin{pmatrix}(\mu+\nu{\mu}^{-1}\nu)_{IJ}&-(\nu{\mu}^{-1})^J_I\\-({\mu}^{-1}\nu)^I_J&({\mu}^{-1})^{IJ}\end{pmatrix}\begin{pmatrix}d\zeta^J\\d\tilde{\zeta}_J\end{pmatrix}=
e^{-2U}\vert\vert \psi\vert\vert^2=\vert\vert {\Psi}\vert\vert^2,
\label{elsofontos}
\eeq
\noindent
i.e. we get back to the usual notation used in the supergravity literature.
Notice that unlike its usual form the new version
as a norm squared is not explicitly $SL(2,\mathbb R)^{\times 3}\subset Sp(8,\mathbb R)$ invariant.
However, by virtue of Eq.(\ref{calc}) we have another interpretation for this term, which clearly displays its $SL(2,\mathbb R)^{\times 3}$ invariance.
(The expression is actually the canonical quadratic $SL(2,\mathbb R)^{\times 4}$ four-qubit invariant, however, the Ehlers $SL(2,\mathbb R)$ transformations of the form $I\otimes I\otimes I\otimes S$ are not preserving the special form of $\vert \Phi\rangle$.)

Proceeding further let us define the $8\times 8$ unitary matrix
\beq
{\bf U}\equiv\begin{pmatrix}{\cal U}\otimes {\cal U}&0\\0&{\cal U}\otimes {\cal U}\end{pmatrix},\qquad
{\cal U}\equiv  HP=\frac{1}{\sqrt{2}}\begin{pmatrix}i&1\\i&-1\end{pmatrix}, 
\label{unitary}
\eeq
\noindent
where $H$ and $P$ are the {\it Hadamard} (discrete Fourier transform) and {\it phase} gates
 known from Quantum Information Theory
\beq
 H=\frac{1}{\sqrt{2}}\begin{pmatrix}1&1\\1&-1\end{pmatrix},\qquad 
 P=\begin{pmatrix}i&0\\0&1\end{pmatrix}.
\label{hadphase}
\eeq
\noindent
We notice that for $\alpha=0,1,2,3$
\beq
{\cal U}\Sigma_{\alpha}{\cal U}^{\dagger}=\frac{i}{y_{\alpha}}\begin{pmatrix}0&d\overline{z}_{\alpha}\\
-d{z}_{\alpha}&0\end{pmatrix}\qquad dz_{j}=dx_j-idy_j,\qquad
dz_0=(dx_0-w)-idy_0,
\label{eztitt}
\eeq
\noindent
with $j=1,2,3$. After introducing the right invariant one-forms ${e}_{\alpha}=\frac{-i}{y_j}dz_{\alpha}$ on the cosets $[SL(2,{\bf \mathbb R})/SO(2)]_{\alpha}$
we can define
\beq
\hat{e}_{\alpha}\equiv {\cal U}\Sigma_{\alpha}{\cal U}^{\dagger}=\begin{pmatrix}0&\overline{e}_{\alpha}\\{e}_{\alpha}&0\end{pmatrix}.
\label{right}
\eeq
\noindent
Using the unitary matrix of Eq.(\ref{unitary}) we can transform
${\cal P}$ of Eq.(\ref{explicit}) to the form
\beq
\hat{\cal P}={\bf U}{\cal P}{\bf U}^{\dagger}=
\frac{1}{2}\begin{pmatrix}\hat{e}_3\otimes I_2+I_3\otimes\hat{e}_2&g\hat{\overline{\Psi}}+\hat{\Psi}g\\
-g\hat{\Psi}^{\dagger}-\hat{\Psi}^Tg&\hat{e}_1\otimes I_0+I_1\otimes\hat{e}_0\end{pmatrix}.
\label{ujp}
\eeq
\noindent
Here we have introduced $\hat{\Psi}=({\cal U}\otimes {\cal U})\Psi({\cal U}\otimes {\cal U})^T$ answering  the new $4$-qubit state
\beq
\vert \hat{\Psi}\rangle =(H\otimes H\otimes H\otimes H)(P\otimes P\otimes P\otimes P)(M_3\otimes M_2\otimes M_1\otimes M_0)\vert d\zeta\rangle .
\label{fourierstate}
\eeq
\noindent
Notice that this new $4$-qubit state is now on the $GL(2,\mathbb C)^{\times 4}$ 
orbit of the one $\vert d\zeta\rangle$ due to the presence of the matrices ${\cal U}\in U(2)$.
Moreover, $\vert \hat{\Psi}\rangle$ can also be regarded as the {\it discrete Fourier transform} of the one $(PM_3\otimes PM_2\otimes PM_1\otimes PM_0)\vert d\zeta\rangle$ incorporating the important phase factors $e^{i\frac{\pi}{2}}$ via the phase gates.

In order to gain some insight into the structure of $\hat{\cal P}$ 
we define the $4\times 4$ matrix 
\beq
\hat{\Phi}=g{\overline{\hat{\Psi}}}g+\hat{\Psi},
\eeq
\noindent
corresponding to the {\it complex} four-qubit state
\beq
\vert\hat{\Phi}\rangle =(\varepsilon\otimes\varepsilon\otimes \varepsilon\otimes \varepsilon)\vert{\overline{\hat{\Psi}}}\rangle +\vert\hat{\Psi}\rangle.
\label{hatfi}
\eeq
\noindent
Notice that though this state is now complex it is again of special form
since it satisfies the {\it reality condition}
\beq
\overline{\vert\hat{\Phi}\rangle}=(\varepsilon\otimes\varepsilon\otimes\varepsilon\otimes\varepsilon)\vert\hat{\Phi}\rangle. 
\label{real}
\eeq
\noindent
In order to understand the structure of $\vert\hat{\Phi}\rangle$
we write its component state $\vert\hat{\Psi}\rangle$ in a three-qubit-like notation
\beq
\vert\hat{\Psi}\rangle=(I\otimes I\otimes I\otimes{\cal U}M_0)\vert\hat{\psi}\rangle,
\quad \vert\hat{\psi}\rangle=({\cal U}M_3\otimes {\cal U}M_2\otimes {\cal U}M_1\otimes I)\vert d\zeta\rangle
\label{kiiras}
\eeq
\noindent
where again $\hat{\psi}_{i_3i_2i_10}\neq 0$ but $\hat{\psi}_{i_3i_2i_11}=0$ thanks to the structure similar to that of $\zeta_{i_3i_2i_1i_0}$.
Introducing the shorthand 
\beq
\hat{\psi}_{i_3i_2i_1}\equiv\hat{\psi}_{i_3i_2i_10}
\eeq
\noindent
in $4\times 4$ notation we get
\beq
\hat{\Psi}=\frac{i}{\sqrt{2y_0}}\begin{pmatrix}\hat{\psi}_{000}&
\hat{\psi}_{000}&\hat{\psi}_{001}&\hat{\psi}_{001}\\
\hat{\psi}_{010}&\hat{\psi}_{010}&\hat{\psi}_{011}&\hat{\psi}_{011}\\
\hat{\psi}_{100}&\hat{\psi}_{100}&\hat{\psi}_{101}&\hat{\psi}_{101}\\
\hat{\psi}_{110}&\hat{\psi}_{110}&\hat{\psi}_{111}&\hat{\psi}_{111}\end{pmatrix}\label{reszlet}
\eeq
\noindent
i.e. the first and the last two columns are the same.
Now using the special structure of the matrix $U\otimes U\otimes U$
one can verify that the following reality conditions hold
\beq
\hat{\psi}_{111}=-\overline{\hat{\psi}}_{000},\quad
\hat{\psi}_{001}=-\overline{\hat{\psi}}_{110},\quad
\hat{\psi}_{010}=-\overline{\hat{\psi}}_{101},\quad
\hat{\psi}_{100}=-\overline{\hat{\psi}}_{011}.
\label{ujreal}
\eeq
\noindent
As a result of these considerations the matrix $\hat{\Phi}$ takes the following form
\beq
\hat{\Phi}=\begin{pmatrix}\overline{{\cal E}}_0&0&0&\overline{{\cal E}_1}\\
0&\overline{{\cal E}_2}&{\cal E}_3&0\\0&\overline{{\cal E}_3}&{\cal E}_2&0\\{\cal E}_1&0&0&{\cal E}_0\end{pmatrix},
\label{alak}
\eeq
\noindent
with
\beq
{\cal E}_0=2\hat{\Psi}_{1110}=i\sqrt{\frac{2}{y_0}}
\hat{\psi}_{111},\quad
{\cal E}_1=2\hat{\Psi}_{1100}=i\sqrt{\frac{2}{y_0}}\hat{\psi}_{110},
\label{eleje}
\eeq
\noindent
\beq
{\cal E}_2=2\hat{\Psi}_{1010}=i\sqrt{\frac{2}{y_0}}\hat{\psi}_{101},\quad
{\cal E}_3=2\hat{\Psi}_{0110}=i\sqrt{\frac{2}{y_0}}\hat{\psi}_{011}.
\label{szerkezet}
\eeq
\noindent
After using this result in the expression for $\hat{\cal P}$ of 
Eq.(\ref{ujp}) we arrive at the explicit form
\beq
\hat{\cal P}=\frac{1}{2}\begin{pmatrix}
0&e_2&e_3&0&\overline{\cal E}_1&0&0&\overline{{\cal E}}_0\\
\overline{e}_2&0&0&e_3&0&-{\cal E}_3&-\overline{\cal E}_2&0\\
\overline{e}_3&0&0&e_2&0&-{\cal E}_2&-\overline{\cal E}_3&0\\
0&\overline{e}_3&\overline{e}_2&0&{\cal E}_0&0&0&{\cal E}_1\\
-{\cal E}_1&0&0&-\overline{\cal E}_0&0&e_0&e_1&0\\
0&\overline{\cal E}_3&\overline{\cal E}_2&0&\overline{e}_0&0&0&e_1\\
0&{\cal E}_2&{\cal E}_3&0&\overline{e}_1&0&0&e_0\\
-{\cal E}_0&0&0&-\overline{\cal E}_1&0&\overline{e}_1&\overline{e}_0&0 \end{pmatrix}.
\label{hosszu}
\eeq
\noindent
The line element in terms of these {\it complex} quantities is the familiar one of Eq.(\ref{line1}) 
\beq
ds^2_{{\cal M}_3}=\sum_{\alpha=0}^3(\overline{e}_{\alpha}e_{\alpha}-\overline{\cal E}_{\alpha}{\cal E}_{\alpha})=\sum_{\alpha=0}^3\frac{d\overline{z}_{\alpha}z_{\alpha}}{y_{\alpha}^2}-\frac{1}{y_0}\vert\vert \hat{\psi}\vert\vert^2
=\sum_{\alpha=0}^3\frac{d\overline{z}_{\alpha}z_{\alpha}}{y_{\alpha}^2}-\vert\vert \hat{\Psi}\vert\vert^2.
\label{line2}
\eeq
\noindent
Here $\vert\vert \hat{\Psi}\vert\vert^2=\langle \hat{\Psi}\vert \hat{\Psi}\rangle$ is the usual scalar product on ${\mathbb C}^{16}$ with complex conjugation in the first factor.

It is important to realize that our quantities ${\cal E}_{\alpha}$ can be written in the familiar form
\beq
{\cal E}_0=\sqrt{2}e^{\frac{K}{2}-U}X^I({\cal N}_{IJ}d\zeta^J-d\tilde{\zeta}_I),\qquad
{\cal E}_j=2i\sqrt{2}y_je^{-U}f^I_j(\overline{\cal N}_{IJ}d{\zeta}^J-d\tilde{\zeta}_I)
\label{nahalisten}
\eeq
\noindent
in terms of the quantities known from special K\"ahler geometry.
Here
\beq 
f_1^I=e^{\frac{K}{2}}D_1X^I=e^{\frac{K}{2}}({\partial}_1+({\partial}_1K))X^I=e^{\frac{K}{2}}\frac{1}{\overline{z}_1-z_1}
\begin{pmatrix}1\\\overline{z}_1\\z_2\\z_3\end{pmatrix},
\qquad{\rm e.t.c.}
\label{specialgeo}
\eeq
\noindent
with $X^I=(1,z_1,z_2,z_3)^T$, $K=-\log(y_1y_2y_3)$ and ${\cal N}_{IJ}$ is defined by Eqs.(\ref{valos}-\ref{kepzetes}).
By virtue of Eqs.(\ref{eleje}-\ref{szerkezet}) these quantities are nicely compressed into the four-qubit state $\vert \hat{\Psi}\rangle$ of Eq.(\ref{fourierstate}).

The special structure of $\hat{\cal P}$ of Eq.(\ref{hosszu}) reveals yet another way for obtaining a four-qubit state. Indeed, $\hat{\cal P}$ contains precisely $16$ nonzero  quantities  which can be organized to form the amplitudes of this new state.
In order to motivate our construction of this new state
let us consider the space of $8\times 8$ matrices of the following form
\beq
\hat{\cal R}=\frac{1}{2}\begin{pmatrix}A&C\\-gC^Tg&B\end{pmatrix},\qquad
C=\begin{pmatrix}0&b&a&0\\
               \overline{d}&0&0&c\\
	      \overline{c}&0&0&d\\
	      0&\overline{a}&\overline{b}&0\end{pmatrix},
\eeq
\noindent
\beq
B=\begin{pmatrix}-\gamma-\delta&0&0&0\\
               0&-\alpha+\beta&0&0\\
              0&0&\alpha-\beta&0\\
              0&0&0&\gamma+\delta\end{pmatrix},\qquad
A=\begin{pmatrix}-\alpha-\beta&0&0&0&\\                                                 0&-\gamma+\delta&0&0\\0&0&\gamma-\delta&0\\
             0&0&0&\alpha +\beta\end{pmatrix}
\label{AB}
\eeq
\noindent
where $\alpha,\beta,\gamma,\delta\in{\mathbb R}$, and $a,b,c,d\in{\mathbb C}$.  Clearly this $12$ real parameter family is  complementary to the $16$ real parameter one characterizing $\hat{\cal P}$. In both cases the off diagonal blocks are related as $C\mapsto -gC^Tg$. In the case of $\hat{\cal P}$ the off-diagonal block satisfies the reality condition $\overline{\hat{\Phi}}=g(\hat{\Phi})g$,
and for $C$ this condition is $\overline{C}=-gCg$. 
Both of the matrices $\hat{\cal P}$ and $\hat{\cal R}$ are satisfying Eq.(\ref{defi}) hence they are elements of the Lie algebra of $SO(4,4,\mathbb C)\simeq SO(8,\mathbb C)$.

Let us now label the rows and columns of these $8\times 8$ matrices 
as $0,1,2,3,4,5,6,7$ or in binary notation $000,001,010,011,100,1010,110,111$.
Now we employ the following permutation to the rows and columns

\beq 
(0,1,2,3,4,5,6,7)\mapsto (7,1,2,4,3,5,6,0),
\eeq
\noindent
\beq
(000,001,010,011,100,101,110,111)\mapsto (111,001,010,100,011,101,110,000).
\label{permutation}
\eeq
\noindent
The binary notation is instructive since it clearly shows that after applying the permutation we get two $4$ element blocks labelled by numbers containing an even number of zeros for the first block and an odd  number of zeros in the second. 
(Another mnemonic: the numbers $1,2,4$ are the quadratic residues modulo 7 and the ones $3,5,6$
are the quadratic nonresidues.)
Now it is easy to check that our fundamental matrix $G$ 
of Eq. (\ref{GG}) is invariant under this permutation.

Applying this permutation to the matrix $\hat{\cal R}$ yields the one
\beq
{\cal R}^{\prime}=\frac{1}{2}\begin{pmatrix}\gamma+\delta&\overline{c}&\overline{d}&0&0&0&0&0\\{c}&-\gamma+\delta&0&\overline{d}&0&0&0&0\\
{d}&0&\gamma-\delta &\overline{c}&0&0&0&0\\
0&{d}&{c}&-\gamma-\delta &0&0&0&0\\
0&0&0&0&\alpha+\beta &\overline{a}&\overline{b}&0\\
0&0&0&0&{a}&-\alpha+\beta &0&\overline{b}\\
0&0&0&0&b&0&\alpha-\beta &\overline{a}\\
0&0&0&0&0&b&{a}&-\alpha-\beta\end{pmatrix}
\label{rvesszo}
\eeq
\noindent
This matrix contains the two $4\times 4$ blocks in its block diagonal part
\beq
\frac{1}{2}({\bf d}{\boldsymbol{\sigma}}\otimes I+I\otimes {\bf c}{\boldsymbol{\sigma}}),\qquad
{\bf d}=\begin{pmatrix}d_1\\d_2\\\delta\end{pmatrix},\quad
{\bf c}=\begin{pmatrix}c_1\\c_2\\\gamma\end{pmatrix},\quad
\quad d=d_1+id_2,\quad c=c_1+ic_2,
\label{dc}
\eeq
\noindent
\beq
\frac{1}{2}({\bf b}{\boldsymbol{\sigma}}\otimes I+I\otimes {\bf a}{\boldsymbol{\sigma}}),\qquad
{\bf b}=\begin{pmatrix}b_1\\b_2\\\beta\end{pmatrix},\quad
{\bf a}=\begin{pmatrix}a_1\\a_2\\\alpha\end{pmatrix},\quad
\quad b=b_1+ib_2,\quad a=a_1+ia_2.
\label{ba}
\eeq
\noindent

The same permutation acting on $\hat{\cal P}$ results in the new form
\beq
{\cal P}^{\prime}_{\ast}=
\frac{1}{2}\begin{pmatrix}0&\Lambda g\\-{\Lambda}^Tg&0\end{pmatrix},
\label{1v}
\eeq
\noindent
\beq
\Lambda=
\begin {pmatrix}{\Lambda}_{0000}&{\Lambda}_{0001}&{\Lambda}_{0010}&{\Lambda}_{0011}\\
{\Lambda}_{0100}&{\Lambda}_{0101}&{\Lambda}_{0110}&{\Lambda}_{0111}\\
{\Lambda}_{1000}&{\Lambda}_{1001}&{\Lambda}_{1010}&{\Lambda}_{1011}\\
{\Lambda}_{1100}&{\Lambda}_{1101}&{\Lambda}_{1110}&{\Lambda}_{1111}\end{pmatrix}\equiv
\begin{pmatrix}-{\cal E}_0&-{e}_0&-{e}_1&-\overline{\cal E}_1\\{e}_2&\overline{\cal E}_2&{\cal E}_3&\overline{e}_3\\
{e}_3&\overline{\cal E}_3&{\cal E}_2&\overline{e}_2\\-{\cal E}_1&-\overline{e}_1&-\overline{e}_0&-\overline{\cal E}_0\end{pmatrix}.
\label{ittalenyeg}
\eeq
\noindent

Now we define a new four-qubit state
\beq
\vert\Lambda\rangle =\sum_{a_3,a_2,a_1,a_0=0,1}{\Lambda}_{a_3a_2a_1a_0}\vert a_3a_2a_1a_0\rangle.
\label{lambdastate}
\eeq
\noindent

Looking at the structure of our matrix ${\cal R}^{\prime}$ it is clear
that it defines an infinitesimal $SU(2)^{\otimes 4}$ action on our state based on the $4\times 4$ {\it complex} matrix $\Lambda$ related to the decomposition

\beq
so(8,{\mathbb C})=[sl(2,\mathbb C)]^4\oplus (2,2,2,2),
\eeq
\noindent
and the embedding of $su(2)$ in $sl(2,\mathbb C)$.

It is important to realize that after the transformation
\beq
C\mapsto C^{\prime}\equiv\begin{pmatrix}
0&-b&-a&0\\
\overline{d}&0&0&-c\\
\overline{c}&0&0&-d\\
0&\overline{a}&\overline{b}&0\end{pmatrix},
\label{cvesszo}
\eeq
\noindent
with $C^{\prime}$ having the property $\overline{C}^{\prime}=gC^{\prime}g$
and also the one of Eq.(\ref{permutation}) the matrix
replacing Eq.(\ref{rvesszo}) will contain the diagonal blocks
	
\beq
\frac{1}{2}({\bf d}{\boldsymbol{\tau}}\otimes I+I\otimes {\bf c}{\boldsymbol{
\tau}}),\qquad
\frac{1}{2}({\bf b}{\boldsymbol{\tau}}\otimes I+I\otimes {\bf a}{\boldsymbol{
\tau}}).\qquad
\label{udc}
\eeq
\noindent
Here the matrices
\beq
{\tau}_1\equiv\frac{i}{2}{\sigma}_2,\qquad
{\tau}_2\equiv -\frac{i}{2}{\sigma}_1,\qquad
{\tau}_3\equiv\frac{1}{2}{\sigma}_3,
\label{su11}
\eeq
\noindent
satisfy the commutation relations of an $su(1,1)$ subalgebra of $sl(2,\mathbb C)$.

Recall that one of our qubits (i.e. the first one labelled by $a_0$)            is still special. According to Eq.(\ref{ittalenyeg}) transformations of the form $I\otimes I\otimes I\otimes S, \quad S=e^{{\frac{i}{2}{\bf a}{\boldsymbol{\sigma}}}}\in SL(2,\mathbb C)$
acting on this qubit relate the first column with the second, and the third with the fourth.
Hence these transformations relate ${\cal E}_0$ to $e_0$ and the  $\overline{\cal E}_j$ to the $e_j$ {\it with the same index} $j$.

Notice that the $SU(2)$ subgroup of this $SL(2,\mathbb C)$ is just the R-symmetry group arising from the restricted holonomy group, of the quaternionic-K\"ahler space which is the analytically continued version of our para-quaternionic  ${\cal M}_3$.
This holonomy implies\cite{Pioline1,Bossard,Feri} that the complexified tangent bundle of that space  splits locally as
${\cal W}\otimes {\cal V}$
where ${\cal W}$ and ${\cal V}$ are vector bundles of dimension $8$ and $2$.
In our case a similar split exists where the former space corresponds to the three-qubit part of our $\Lambda$ labelled by the indices $a_3, a_2, a_1$ and the latter to the special qubit labelled by $a_0$.
Indeed our transformations of Eqs.(\ref{unitary}) and (\ref{permutation}) correspond to a change of basis in $T_{\mathbb C}{\cal M}_3$ similar to the usual one rendering the "quaternionic vierbein" covariantly constant with respect to the spin connection\cite{Bossard}.

Eqs.(\ref{1v}-\ref{lambdastate}) are of central importance for our considerations of the following sections.
They define a {\it complex} four-qubit state satisfying the reality condition 
\beq
\overline{\vert\Lambda\rangle} =(\sigma_1\otimes \sigma_1\otimes \sigma_1\otimes \sigma_1)\vert\Lambda\rangle,\qquad \sigma_1=\begin{pmatrix}0&1\\1&0\end{pmatrix},
\label{realagain}
\eeq
\noindent
where $\sigma_1$ is the bit flip gate of Quantum Information Theory.
It is straightforward to check that the subgroup of transformations of the group $SL(2,\mathbb C)^{\times 4}$ leaving invariant this reality condition is $SU(1,1)^{\times 4}$ i.e. precisely those transformations as described by Eqs.(\ref{cvesszo})-(\ref{su11}).
Hence in the notation used in quantum information the admissible transformations are of the form
\beq
\vert\Lambda\rangle\mapsto (S\otimes S_2\otimes S_1\otimes S_0)\vert\Lambda\rangle,\qquad S_3,S_2,S_1,S_0\in SU(1,1).
\eeq
\noindent

In what follows our basic concern will be a study of quantities invariant under the larger group of  transformations i.e. $SL(2,\mathbb C)^{\times 4}$. 
Such invariants are clearly also $SU(1,1)^{\times 4}$ ones.
It is known that the number of such algebraically independent invariants is four\cite{Luque,Osterloh}. We have a quadratic, two quartic, and one sextic invariant.
The structure and geometry of such invariants has been investigated\cite{Levay41,Osterloh}.
Here in closing this section we just observe that the quadratic four-qubit invariant\cite{Luque} for our state $\vert\Lambda\rangle$ is precisely the line element $ds^2_{{\cal M}_3}$ i.e.
\beq
ds^2_{{\cal M}_3}=-\frac{1}{2}{\varepsilon}^{a_3a_3^{\prime}}
{\varepsilon}^{a_2a_2^{\prime}}{\varepsilon}^{a_1a_1^{\prime}}{\varepsilon}^{a_0a_0^{\prime}}{\Lambda}_{a_3a_2a_1a_0}{\Lambda}_{a_3^{\prime}a_2^{\prime}a_1^{\prime}a_0^{\prime}}=\sum_{\alpha=0}^3(\overline{e}_{\alpha}e_{\alpha}-\overline{\cal E}_{\alpha}{\cal E}_{\alpha}).
\label{lineasinv}
\eeq
\noindent
This formula first appeared in the paper of Bossard et.al.\cite{Bossard}
Here we have also clarified its intimate connection to four-qubit systems.
We also remark that the quadratic invariant is also a permutation invariant.
However, from the physical point of view the special 
role we have attached to the first qubit obviously breaks this permutation invariance.

\section{Conserved charges}

The $3D$ duality group acts isometrically on our ${\cal M}_3$
by right multiplication and yields a conserved Noether charge\cite{Bossard,Stelle,Trigiante} 
\beq
Q=V^{-1}{\cal P}V=\begin{pmatrix}Q_{11}&-gQ_{12}\\gQ_{12}^T&Q_{22}\end{pmatrix}.
\label{Noether}
\eeq
\noindent
The explicit expression of $Q$ is given by
\beq
2Q=
\begin{pmatrix}{\bf 1}&\zeta g\\-\zeta^Tg&{\bf 1}+\frac{1}{2}\Delta\end{pmatrix}
\begin{pmatrix}\rho_3\otimes I_2+I_3\otimes \rho_2&-gd\hat{\zeta}-d\zeta g\\
gd\hat{\zeta}^T+d\zeta^Tg&\rho_1\otimes I_0+I_1\otimes\rho_0\end{pmatrix}
\begin{pmatrix}{\bf 1}&-\zeta g\\\zeta^Tg&{\bf 1}+\frac{1}{2}\Delta\end{pmatrix}
\label{explnoether}
\eeq
\noindent
where
\beq
\rho_{\alpha}\varepsilon\equiv {\rm Re}\left(\frac{dz_{\alpha}}{y_{\alpha}^2}\begin{pmatrix}\overline{z}_{\alpha}\\1\end{pmatrix}\begin{pmatrix}\overline{z}_{\alpha}&1\end{pmatrix}\right),
\quad z_{\alpha}=x_{\alpha}-iy_{\alpha},\quad \alpha=0,1,2,3
\label{rho}
\eeq
\noindent
and

\beq
d\hat{\zeta}\equiv (N_3\otimes N_2)d\zeta (N_1\otimes N_0),\qquad N_{\alpha}\equiv M_{\alpha}^TM_{\alpha},\quad \alpha=0,1,2,3
\label{N}
\eeq
\noindent
where for $dz_j, j=1,2,3$ and $dz_0$ we used the definitions of Eq.(\ref{eztitt}).

Now we are interested in the conserved electric and magnetic charges
coming from the first and third column of $Q_{12}$. (This part has the same structure as the matrix $\zeta$ of Eq.(\ref{zetaparam})). Since the matrix $\zeta g$
has vanishing first and third column, any $4\times 4$ matrix multiplied by $\zeta g$ from the {\it right} also has this property. Hence terms having this structure will not contribute to the relevant part of $Q_{12}$.
Using $\Delta=-\zeta^Tg\zeta g$ the relevant part of $Q_{12}$ is
\beq
2[Q_{12}]_{\rm relevant}=\frac{1}{y_0}(N_3\otimes N_2)d\zeta(N_1\otimes I)+\frac{dx_0-w}{y_0^2}(\varepsilon\otimes\varepsilon)\zeta(\varepsilon\otimes I).
\label{relevans}
\eeq
\noindent
On the other hand let us look at the conserved quantity
\beq
k\equiv\frac{1}{2}{\rm Tr}(I_1\otimes E_0)Q_{22}=\frac{(dx_0-w)}{2y_0^2}.
\label{nut}
\eeq
\noindent
Since the line element is related to the Lagrangian and the Lagrangian according to Eq.(\ref{line1}) contains a term $((\dot{x}_0-w)^2+\dot{y_0}^2)/4y_0^2$ hence $p_{x_0}=p_{\sigma}=\frac{\partial L}{\partial{\dot{x}_0}}=k$ i.e. our quantity is just the NUT charge.
These considerations show that the relevant part $\Gamma$ of $Q_{12}$ in four-qubit notation is
\beq
\vert \Gamma\rangle =\frac{1}{2y_0}(N_3\otimes N_2\otimes N_1\otimes I )\vert d\zeta\rangle
-p_{x_0}(\varepsilon\otimes\varepsilon\otimes\varepsilon\otimes I)\vert\zeta\rangle.
\label{q12qubit}
\eeq
\noindent
Eq.(\ref{q12qubit}) comprises $8$ conserved quantities represented as the $8$ nonzero amplitudes of a four-qubit state.
Note, that our formula also contains the NUT charge.
In a three-qubit-like notation we can alternatively write this as
\beq
\vert \Gamma\rangle =\frac{1}{2}e^{-2U}({\cal N}\otimes I)\vert d\zeta\rangle -{p_{\sigma}}
(\epsilon\otimes I)\vert\zeta\rangle,
\label{ezagamma}
\eeq
\noindent
with 
\beq
{\cal N}\equiv N_3\otimes N_2\otimes N_1,\qquad \epsilon\equiv \varepsilon\otimes\varepsilon\otimes\varepsilon.
\eeq
\noindent

Let us see how this set of conserved quantities is related to the momenta $p_{{\zeta}^I}=\frac{\partial L}{\partial\dot{\zeta}^{I}}$
and $p_{\tilde{\zeta}_I}=\frac{\partial L}{\partial\dot{\tilde{\zeta}}_{I}}$.
A calculation based on the Lagrangian related to the line element Eq.(\ref{line1}) shows that these quantities can be also organized into a state
\beq
\vert p_{\zeta}\rangle\equiv-\frac{1}{2}e^{-2U}({\cal N}\otimes I)\vert d{\zeta}\rangle+\frac{p_{\sigma}}{2}(\epsilon\otimes I)\vert\zeta\rangle.
\label{conjmomenta}
\eeq
\noindent
Hence
\beq
\vert \Gamma\rangle =-\frac{p_{\sigma}}{2}(\epsilon\otimes I)\vert\zeta\rangle -\vert p_{\zeta}\rangle.
\label{osszefugg}
\eeq
\noindent

Let us now introduce the new quantity
\beq
\vert P_{\zeta}\rangle\equiv \vert p_{\zeta}\rangle -\frac{p_{\sigma}}{2}(\epsilon\otimes I)\vert\zeta\rangle.
\label{bigp}
\eeq
\noindent
Notice that after writing out the $8$ amplitudes explicitly we get
\beq
P_{{\zeta}^I}=p_{{\zeta}^I}-\frac{p_{\sigma}}{2}\tilde{\zeta}_I,\qquad
P_{\tilde{\zeta}_I}=p_{\tilde{\zeta}_I}+\frac{p_{\sigma}}{2}{\zeta}^I,
\label{accord}
\eeq
\noindent
in accordance with Eq. (4.14) of Bossard et.al.\cite{Bossard} (The $\sigma$ used by them is different by a factor of $2$).
Now one can verify that the Hamiltonian is
\beq
H=\sum_{\alpha =0}^3y_{\alpha}^2(p_{x_{\alpha}}^2+p_{y_{\alpha}}^2)-
y_0\langle P_{\zeta}\vert {\cal N}^{-1}\otimes I
\vert P_{\zeta}\rangle.
\label{Hamiltonian}
\eeq
\noindent
From Eqs.(\ref{osszefugg}-\ref{bigp}) we have
\beq
\vert\hat{\Gamma}\rangle\equiv\vert\Gamma\rangle +
{p_{\sigma}}(\epsilon\otimes I)\vert\zeta\rangle=-                     \vert P_{\zeta}\rangle 
\label{bigpexpl}
\eeq
\noindent
arriving at an alternative expression as

\beq
H=\sum_{\alpha =0}^3y_{\alpha}^2(p_{x_{\alpha}}^2+p_{y_{\alpha}}^2)-
e^{2U}\langle \hat{\Gamma}\vert {\cal N}^{-1}\otimes
 I\vert \hat{\Gamma}\rangle
\label{3hami}
\eeq
\noindent
in a three-qubit-like notation.

Let us now parametrize $\Gamma$ in terms of the electric and magnetic charges as
\beq
\Gamma =\frac{1}{\sqrt{2}}\begin{pmatrix}p^0&0&-p^1&0\\
-p^2&0&q_3&0\\-p^3&0&q_2&0\\q_1&0&q_0&0\end{pmatrix},
\label{electricmagnetic}
\eeq
\noindent
where the rows and columns of this matrix are related to the four-qubit labels  as in Eq.(\ref{zetaparam}) defining the four-qubit state $\vert\Gamma\rangle$.
This parametrization in the conventional language amounts to 
\beq
\frac{1}{\sqrt{2}}p^I=\frac{1}{2}p_{\sigma}\zeta^I+p_{\tilde{\zeta}_I},\qquad
\frac{1}{\sqrt{2}}q^I=\frac{1}{2}p_{\sigma}\tilde{\zeta}_I-p_{{\zeta}^I}.
\label{conv3}
\eeq
\noindent
Notice that the origin of the factors of $\sqrt{2}$s appearing in Eqs.(\ref{electricmagnetic}-\ref{conv3}) can be traced back to the fact\cite{Trigiante,Bossard} that the electric and magnetic charges should be proportional to the generators $\sqrt{2}E_{p^i}$ and $\sqrt{2}E_{q_I}$.
For vanishing NUT charge $p_{\sigma}=0$ we get
\beq
e^{2U}V_{BH}
=e^{2U}\langle\Gamma\vert {\cal N}^{-1}\otimes
I\vert\Gamma\rangle
\label{BHpot}
\eeq
\noindent
which can be checked to yield the usual expression for the $V_{BH}$ black hole potential.

Let us now rewrite our expression of Eq.(\ref{state}) for $\vert\Psi\rangle$ in terms of our conserved quantities.
First by using Eqs.(\ref{ezagamma}) and (\ref{bigpexpl}) we express $\vert d\zeta\rangle$ in terms of the charges as
\beq
\vert d\zeta\rangle =2e^{2U}({\cal N}^{-1}\otimes I)\vert\hat{\Gamma}\rangle,
\label{kifejez}
\eeq
\noindent
to arrive at the expression
\beq
\vert\Psi\rangle=2e^{U}(M_3^T\otimes M_2^T\otimes M_1^T\otimes I)^{-1}\vert\hat{\Gamma}\rangle.
\label{ittaveg}
\eeq
\noindent
We can further transform this to obtain $\vert\hat{\Psi}\rangle=({\cal U}\otimes {\cal U}\otimes {\cal U}\otimes {\cal U})\vert\Psi\rangle$ of Eq.(\ref{fourierstate})
\beq
\vert\hat{\Psi}\rangle=2e^{2U}(I\otimes I\otimes I\otimes {\cal U}M_0)({\cal V}\otimes{\cal V}\otimes {\cal V}\otimes I)(S_3\otimes S_2\otimes S_1\otimes I)\vert\hat{\gamma}\rangle,
\label{hosszu2}
\eeq
\noindent
where we have used that
\beq
{\cal U}{M^T}^{-1}={\cal V}S\sigma_3.
\eeq
\noindent
Here the matrices ${\cal V}$ and $S_j$, $j=1,2,3$ are the ones of Eq.(\ref{Smatrix})
discussed in the Introduction and
\beq
\vert\hat{\gamma}\rangle =(\sigma_3\otimes \sigma_3\otimes\sigma_3\otimes I)\vert\hat{\Gamma}\rangle.
\label{gammahat}
\eeq
\noindent
Clearly for $k=0$ i.e. vanishing NUT charge $\vert\gamma\rangle$ is just the phase-flipped version of $\vert\Gamma\rangle$ of Eq.(\ref{electricmagnetic}).
This $\vert\gamma\rangle$ reinterpreted as a three-qubit state is just the charge state mentioned in Eq.(\ref{gammacharge}). 
Now notice that we have 
\beq
{\cal U}M_0=\frac{i}{\sqrt{2}}e^{-U}\begin{pmatrix}1&\ast&\\1&\ast\end{pmatrix},
\eeq
\noindent
where the terms in the second column are not needed since for the four-qubit state  $\vert\hat{\Gamma}\rangle$ we have as usual  $\hat{\Gamma}_{j_3j_2j_11}=0$,
a structure that dates back to the similar one of $\vert\zeta\rangle$ and $\vert\Gamma\rangle$.
Now obviously $\hat{\Psi}_{j_3j_2j_11}= 
\hat{\Psi}_{j_3j_2j_10}$ and either of them can be reinterpreted as the ones of a
three-qubit
 state. (See also Eq. (\ref{reszlet}) in this respect.) For this three-qubit projection we have
\beq
\vert\hat{\Psi}\rangle_3\equiv i\sqrt{2}\vert\hat{\chi}\rangle =i\sqrt{2}e^{U}({\cal V}\otimes {\cal V}    \otimes 
{\cal V})(S_3\otimes S_2\otimes S_1)\vert\hat{\gamma}\rangle,
\label{ezittaklassz}
\eeq
\noindent
 where by virtue of Eqs. (\ref{gammahat}) and (\ref{bigpexpl})
\beq
\vert\hat{\gamma}\rangle=\vert\gamma\rangle+{p_{\sigma}}(\sigma_1\otimes\sigma_1\otimes\sigma_1)\vert\zeta\rangle.
\label{naeztkapdki}
\eeq
\noindent

Eqs. (\ref{ezittaklassz}) and (\ref{naeztkapdki}) clearly show how our state $\vert\chi\rangle$ of Eq. (\ref{chi})  
as a special case of $\vert\hat{\chi}\rangle$
is embedded in a four-qubit state $\vert\hat{\Psi}\rangle$.
Moreover, in achieving this we managed to present a generalization also valid in the case of nonvanishing NUT charge.
The important new property to be noted here is that unlike $\vert\gamma\rangle$
which is constant the one $\vert\hat{\gamma}\rangle$ is depending on $\tau=\frac{1}{r}$ via the Wilson lines $\zeta^I$ and $\tilde{\zeta}_I$.
Notice that we also have $\vert\vert\hat\Psi\vert\vert^2=4e^{2U}V_{BH}=4\vert\vert\hat{\chi}\vert\vert^2_3$, i.e.
the Black Hole potential is just the norm of a three-qubit state. \cite{Levay3,Levszal}.

For later use for static spherically symmetric solutions let us write out eplicitly the quantities ${\cal E}_{\alpha}$ of Eq.(\ref{eleje})-(\ref{szerkezet}) in terms of the three-qubit state $\vert\chi\rangle$ of Eqs.(\ref{chi}) and (\ref{3qubitallapot})
\beq
{\cal E}_0=i\sqrt{8}{\chi}_{111},\quad
{\cal E}_1=i\sqrt{8}{\chi}_{110},\quad
{\cal E}_2=i\sqrt{8}{\chi}_{101},\quad
{\cal E}_3=i\sqrt{8}{\chi}_{011},\quad
\label{1epi}
\eeq
\noindent
\beq                                                                            \overline{\cal E}_0=i\sqrt{8}{\chi}_{000},\quad
\overline{\cal E}_1=i\sqrt{8}{\chi}_{001},\quad
\overline{\cal E}_2=i\sqrt{8}{\chi}_{010},\quad
\overline{\cal E}_3=i\sqrt{8}{\chi}_{100}.\quad
\label{2epi}
\eeq
\noindent
For the more general stationary solutions we have to use $\vert\hat{\chi}\rangle$ as given by Eq.(\ref{ezittaklassz}).

In closing this section let us also calculate the conserved quantity
\beq
m\equiv\frac{1}{4}{\rm Tr}(I_1\otimes H_0)Q_{22}.
\label{ADM}
\eeq
\noindent
Writing out explicitly $Q_{22}$ using Eq.(\ref{explnoether})
we notice that many terms end with the matrix $\zeta g$. Using the cyclic property of the trace these terms in some cases result in ones begining 
with $\zeta g(I_1\otimes H_0)\zeta^Tg$ which is vanishing.
Employing Eq.(\ref{conv3}) and the definition $p_{y_0}=\frac{\dot{y}_0}{2y_0^2}$ the result of these considerations will be just two nonvanishing terms yielding the final result
\beq
m=\frac{1}{2}\langle\Gamma\vert\zeta\rangle+x_0p_{x_0}+{y_0}p_{y_0}=-\frac{1}{2}(\zeta^Ip_{\zeta^I}+\tilde{\zeta}_Ip_{\tilde{\zeta}_I})+\sigma p_{\sigma}+\dot{U}.
\label{ADM2}
\eeq
\noindent
which is the ADM mass of the black hole\cite{Stelle,Stelle2,pioline2}.

\section{Black hole solutions as entangled systems}
\subsection{BPS solutions}

Let us consider our four-qubit state $\vert\Lambda\rangle$ of Eqs.(\ref{1v})-(\ref{lambdastate}).
In this section we would like to investigate issues of separability for this state.
In particular in this subsection we will be interested in the sufficient and necessary condition for the separability of the {\it first qubit}, i.e. the one which is labelled by $a_0$ in Eq.(\ref{lambdastate}).
From our previous considerations it is clear that this qubit is the one of special status, i.e. it is the one transforming as a doublet under the $R$-symmetry.

In QIT terms separability of this qubit from the rest is {\it equivalent} to the condition that the (unnormalized) $2\times 2$ reduced density matrix ${\varrho}_1\equiv{\rm Tr}_1\vert \Lambda\rangle\langle\Lambda\vert$  
represents a pure state\cite{Beng}.
This density matrix is of the form
\beq
{\varrho}_1=\begin{pmatrix}\langle\Lambda_0\vert\Lambda_0\rangle&
\langle\Lambda_0\vert\Lambda_1\rangle\\ \langle\Lambda_1\vert\Lambda_0\rangle&\langle\Lambda_1\vert\Lambda_1\rangle\end{pmatrix},\qquad
\langle\Lambda_{a_0}\vert\Lambda_{a^{\prime}_0}\rangle\equiv \sum_{a_3,a_2,a_1=0,1}
\overline{\Lambda}_{a_3a_2a_1a_0}\Lambda_{a_3a_2a_1a^{\prime}_0}.
\eeq
\noindent
This is a pure state i.e. a projector if and only if ${\rm Det}\varrho_1=0$.
Equivalently this condition is satisfied iff 
the two $8$ component vectors $\Lambda_{a_3a_2a_10}$ and ${\Lambda}_{a_3a_2a_11}$
are proportional, i.e.
$\Lambda_{a_3a_2a_10}=\lambda{\Lambda}_{a_3a_2a_11}$.
By virtue of the reality condition of Eq.(\ref{realagain}) we  also have the constraint $\vert\lambda\vert =1$. 
Using the definitions in Eq.(\ref{ittalenyeg}) this means that
\beq
{\cal E}_0=\lambda e_0,\qquad {\cal E}_j=\lambda\overline{e}_j,\qquad 
\vert\lambda\vert=1.
\label{bps}
\eeq
\noindent
The first consequence of these considerations is that for the state $\Lambda$ the quadratic invariant $I_1$ of the Appendix 
(see Eq.(\ref{26})) which is related to $ds^2$ of Eq.(\ref{lineasinv})  is vanishing.
Moreover, since the first column of the matrix $\Lambda$ of Eq.(\ref{ittalenyeg}) is proportional to the second and the third one is proportional to the fourth, the invariant $I_4$ which is according to Eq.(\ref{detinv}) just the derminant of ${\Lambda}$ is also vanishing.
A straightforward calculation based on Eqs.(\ref{sexkiirva}) and (\ref{last4})
shows that the remaining two algebraically independent invariants $I_3$ and $I_2$ are also vanishing.

Now the $4\times 4$ matrix $\Omega\equiv\Lambda^Tg\Lambda g$ satisfies Eq.(\ref{poli}) of the Appendix,
so it follows that ${\Omega}^4=0$.
This implies that the matrix ${\cal R}_{\Lambda}$ of Eq.(\ref{rmatrix}) is nilpotent.
According to the terminology of Ref.\cite{Djokovic}
states with the property that their  associated $8\times 8$ matrix ${\cal R}_{\Lambda}$ is nilpotent are called {\it nilpotent states}. 
Hence our separable state is a (trivial) example of a nilpotent $4$-qubit state.
Moreover, since ${\cal R}_{\Lambda}$ is just $2{\cal P}^{\prime}$ of Eq.(\ref{1v})
and this matrix is unitarily related to the matrix $Q$ of Eq.(\ref{Noether}) of conserved charges, it follows that $Q$ is also nilpotent. 

In order to link these considerations to the usual static, extremal spherically symmetric BPS black hole solutions
we choose $\lambda$ as
\beq
\lambda=-i\sqrt{\frac{Z}{\overline{Z}}}.
\label{bps2}
\eeq
\noindent
Note that for static solutions the NUT charge is zero, hence $x_0=0$ and $e_0=-\frac{dy_0}{y_0}=d\phi$, i.e. $\overline{e_0}=e_0$. 

Now in the language of supergravity the above discussed condition on separability is just the usual one on the existence of Killing spinors\cite{pioline2,Bossard}
expressed in terms of the quaternionic vierbein
\beq
{\Lambda}_{a_3a_2a_1a_0}{\epsilon}^{a_0}=0,\qquad \epsilon^{a_0}=\begin{pmatrix}1\\\lambda\end{pmatrix},
\label{quatvier}
\eeq
\noindent
and Eqs.(\ref{bps}), (\ref{nahalisten}) and (\ref{central}),  give rise to the attractor flow equations \cite{attractor,Bossard}
\beq
\dot{U}=-e^{U}\vert Z\vert,\qquad \dot{z}^j=-2e^{U}G^{j\overline{k}}{\partial}_{\overline{k}}\vert Z\vert.
\label{bpsexpl}
\eeq
\noindent
As it is well-known these first order equations imply that the corresponding second order equations of Eq.(\ref{Euler}) hold.
Moreover, by virtue of the vanishing of the invariant $I_1$ the constraint of Eq.(\ref{constraint}) is also satisfied hence the solution is extremal.

From this analysis we have learnt that the condition of separability for the first qubit for the $4$-qubit state $\vert\Lambda\rangle$ taken together with the special choice of Eq.(\ref{bps2}) yields the first order attractor flow equations.
Moreover, we have seen that in this case $\vert\Lambda\rangle$ is a {\it nilpotent} state. This property of $\vert\Lambda\rangle$ is related to the well-known nilpotency of the Noether charge\cite{Bossard,Trigiante,Stelle} $Q$.
Notice however, that our approach does not directly yield the order of nilpotency of $Q$ for BPS solutions which is three\cite{Bossard}.

\subsection{Non-BPS solutions with vanishing central charge}

Let us discuss the separability properties of $\vert\Lambda\rangle$
associated with the remaining qubits not playing any distinguished role.
Here we chose to consider separability of the $4$th qubit.
An argument similar to the one as given in the previous subsection shows
that the sufficient and necessary condition of separability for this qubit is
that the first row is proportional to the third and the second is proportional to the fourth. Due to the reality condition we again have $\vert\lambda\vert=1$ and we get
\beq
{\cal E}_0=-\lambda e_3,\qquad \overline{\cal E}_1=-\lambda\overline{e}_2,
\qquad \overline{\cal E}_2=-\lambda\overline{e}_2,\qquad {\cal E}_3= -\lambda\overline{e_0}.
\label{furafelt}
\eeq
\noindent
Using the definitions of Eq.(\ref{nahalisten}) these conditions take the explicit form
\beq
\frac{\dot{z}_0}{y_0}=\overline{\lambda}e^UZ_3,\qquad
\frac{\dot{z}_1}{y_1}=\lambda e^UZ_2,\qquad\frac{\dot{z}_2}{y_2}=\lambda e^UZ_1,\qquad \frac{\dot{z}_3}{y_3}=-\overline{\lambda}e^U Z,
\label{veryexpl}
\eeq
\noindent
where $Z_j\equiv -2iy_jD_jZ$ with $D_j$ as given by Eq.(\ref{kovika}).
Now for static solutions we again have no twist potential i.e. $x_0=0$ hence by choosing
\beq
\lambda =-i\sqrt{\frac{Z_3}{\overline{Z_3}}}
\label{lambdika}
\eeq
\noindent
we get
\beq
\dot{U}=-e^U\vert Z_3\vert,\qquad \dot{z}^j=-2e^UG^{j\overline{k}}{\partial}_{\overline{k}}\vert Z_3\vert.
\label{vanish4}
\eeq
\noindent
These expressions show that demanding separability for the fourth qubit taken together with the choice of Eq.(\ref{lambdika}) yields the first order equations characterizing attractors with vanishing central charge\cite{Scherbakov}.

Clearly similar considerations apply for issues of separability for the second and third qubits. The result will be similar sets of equations with
$\vert Z_3\vert$ replaced by $\vert Z_1\vert$ and $\vert Z_2\vert$.
This amounts to taking different forms for the fake superpotential\cite{Bossard}.

Note that the value for the four-qubit invariant $I_1$ is related to the extremality parameter.
Unlike the other three algebraically independent invariants, this is also a permutation invariant. 
Of course the value of $I_1$ is zero for both BPS and non-BPS solutions with vanishing central charge, expressing the fact that our solutions are extremal.
Moreover, for all of our non-BPS solutions some rows or columns of the          $4\times 4$ matrix are proportional, hence the invariant $I_4$ is zero as well.
Calculations show that the remaining invariants $I_2$ and $I_3$ also give zero, hence our considerations on the nilpotency of $\vert\Lambda\rangle$ familiar from the previous subsection still apply.

In closing this subsection we note that the conditions for separability can be written in the familiar form\cite{Bossard} of Eq.(\ref{quatvier}) with the label of ${\epsilon}^{a_{\alpha}}$ is ${a_0}$ for BPS, $a_j$, $j=1,2,3$ for non-BPS solutions with vanishing central charge.
Of course  $\lambda$ should be modified accordingly.

\subsection{Non-BPS seed solutions}

From the previous subsections it is obvious that the condition of extremality
related to the vanishing of the invariant $I_1$ can be satisfied in a number of different ways. Explicitly the relevant equation to be satisfied is
\beq
\sum_{\alpha=0}^3\overline{\cal E}_{\alpha}{\cal E}_{\alpha}=\sum_{\alpha=0}^3  \overline{e}_{\alpha}e_{\alpha}.
\label{hmm}
\eeq
\noindent
For static solutions we have already remarked that $\overline{e}_0=e_0$, hence
for BPS solutions Eqs. (\ref{bps})-(\ref{bps2})
can be written in the form ${\cal E}_{\alpha}=\lambda\overline{e}_{\alpha}$,
i.e. ${\cal E}_{\alpha}$ is related to $\overline{e}_{\alpha}$ 
via a special element of $U(4)$ containing merely phase factors $\lambda$ in  its diagonal.
In the case of non-BPS solutions with vanishing central charge Eqs.(\ref{furafelt})-(\ref{lambdika}) of the previous subsection can be written in a similar way in terms of another element of $U(4)$ 
\beq
\begin{pmatrix}{\cal E}_0\\{\cal E}_1\\{\cal E}_2\\{\cal E}_3\end{pmatrix}=
\begin{pmatrix}0&0&0&-\lambda\\0&0&-\overline{\lambda}&0\\
0&-\overline{\lambda}&0&0\\-\lambda&0&0&0\end{pmatrix}\begin{pmatrix}e_0\\e_1\\e_2\\e_3\end{pmatrix}.
\eeq
\noindent
Similarly the basic equations of the remaining two cases of the previous subsection can be expressed in terms of similar unitaries.
These unitaries are just permutation matrices combined with phase factors and their conjugates. As we have shown this structure is related to the separability of one of the qubits of the state $\vert\Lambda\rangle$.
In simple terms this means that some of the rows or columns of the $4\times 4$ matrix $\Lambda$ corresponding to $\vert\Lambda\rangle$ are proportional to each other.

In order to obtain states $\vert\Lambda\rangle$ which are {\it entangled}
and at the same time give rise to static spherically symmetric non-BPS black hole solutions with non-vanishing central charge we have to experiment with elements of $U(4)$ of more general type.

Let us consider the following choice
\beq
\begin{pmatrix}{\cal E}_0\\{\cal E}_1\\{\cal E}_2\\{\cal E}_3\end{pmatrix}=-\frac{i}{2}
\begin{pmatrix}1&1&1&1\\1&1&-1&-1\\
1&-1&1&-1\\1&-1&-1&1\end{pmatrix}\begin{pmatrix}e_0\\e_1\\e
_2\\e_3\end{pmatrix}.
\label{nonbpsunitary}
\eeq
\noindent
Due to the unitarity of the relevant matix the condition of extremality is satisfied, moreover obviously none of the qubits can be separated from the rest.
However, apart from satisfying Eq.(\ref{hmm}) or equivalently Eq.(\ref{constraint})
we still have to satisfy the equations of motion i.e. Eq.(\ref{Euler}). 
In the following we show that the choice of Eq.(\ref{nonbpsunitary}) indeed gives rise to a solution of the latter equations namely
the non-BPS seed solution\cite{seed}.
Clearly apart from characterizing the seed solution in a nice and compact way Eq.(\ref{nonbpsunitary}) also serves as a mnemonic for the structure of the corresponding entangled $4$-qubit state $\vert\Lambda\rangle$ of Eqs.(\ref{1v})-(\ref{lambdastate}).

In order to reveal the structure
of the seed solution for special non-BPS charge configurations 
we recall Eqs.(\ref{1epi})-(\ref{2epi}) and (\ref{chi}) and employ a discrete Fourier (Hadamard) transformation to
$\vert\chi\rangle$ as
\beq
\vert\tilde{\chi}(\tau)\rangle =(H\otimes H\otimes H)\vert\chi(\tau)\rangle,\qquad H=\frac{1}{\sqrt{2}}\begin{pmatrix}1&1\\1&-1\end{pmatrix}.
\label{hadi}
\eeq
\noindent
The amplitudes of this state are
\beq
\sqrt{2y_1y_2y_3}\tilde{\chi}_{000}=-ie^Uy_1y_2y_3p^0,
\label{1x}
\eeq
\noindent
\beq
\sqrt{2y_1y_2y_3}\tilde{\chi}_{110}=ie^Uy_1(x_2x_3p^0-x_2p^3-x_3p^2+q_1)
\label{2x}
\eeq
\noindent
\beq
\sqrt{2y_1y_2y_3}\tilde{\chi}_{101}=ie^Uy_2(x_1x_3p^0-x_1p^3-x_3p^1+q_2),
\label{3x}
\eeq
\noindent
\beq
\sqrt{2y_1y_2y_3}\tilde{\chi}_{011}=ie^Uy_3(x_1x_2p^0-x_1p^2-x_2p^1+q_3),
\label{4x}
\eeq
\noindent
\beq
\sqrt{2y_1y_2y_3}\tilde{\chi}_{111}=e^U(x_1x_2x_3p^0-x_2x_3p^1-x_1x_3p^2-x_1x_2p^3+x_1q_1+x_2q_2+x_3q_3+q_0),
\label{5x}
\eeq
\noindent
\beq
\sqrt{2y_1y_2y_3}\tilde{\chi}_{001}=e^Uy_2y_3(p^1-x_1p^0),
\label{6x}
\eeq
\noindent
\beq
\sqrt{2y_1y_2y_3}\tilde{\chi}_{010}=e^Uy_1y_3(p^2-x_2p^0),
\label{7x}
\eeq
\noindent
\beq
\sqrt{2y_1y_2y_3}\tilde{\chi}_{001}=e^Uy_1y_2(p^3-x_3p^0).
\label{8x}
\eeq
\noindent
Now one can check that Eq.(\ref{nonbpsunitary}) can be expressed in terms of these quantities as
\beq
\tilde{\chi}_{000}=\frac{i}{2}\frac{\dot{x_0}}{y_0},\qquad
\tilde{\chi}_{110}=\frac{i}{2}\frac{\dot{x_1}}{y_1},\qquad
\tilde{\chi}_{101}=\frac{i}{2}\frac{\dot{x_2}}{y_2},\qquad
\tilde{\chi}_{011}=\frac{i}{2}\frac{\dot{x_3}}{y_3},
\label{elsofele}
\eeq
\noindent
\beq
\tilde{\chi}_{111}=\frac{1}{4}\left(\frac{\dot{y_0}}{y_0}-
\frac{\dot{y_1}}{y_1}-
\frac{\dot{y_2}}{y_2}-
\frac{\dot{y_3}}{y_3}\right),\qquad
\tilde{\chi}_{001}=\frac{1}{4}\left(-\frac{\dot{y_0}}{y_0}+
\frac{\dot{y_1}}{y_1}-
\frac{\dot{y_2}}{y_2}-
\frac{\dot{y_3}}{y_3}\right)
\label{masodikfele1}
\eeq
\noindent
\beq
\tilde{\chi}_{010}=\frac{1}{4}\left(-\frac{\dot{y_0}}{y_0}-
\frac{\dot{y_1}}{y_1}+
\frac{\dot{y_2}}{y_2}-
\frac{\dot{y_3}}{y_3}\right),\qquad
\tilde{\chi}_{100}=\frac{1}{4}\left(-\frac{\dot{y_0}}{y_0}-
\frac{\dot{y_1}}{y_1}-
\frac{\dot{y_2}}{y_2}+
\frac{\dot{y_3}}{y_3}\right).
\label{masodikfele2}
\eeq
\noindent
For static solutions we have vanishing NUT charge i.e. $x_0=0$ hence the first of these equations reads as $\tilde{\chi}_{000}=0$ which by virtue of Eq.(\ref{1x})means that $p^0=0$. 
Hence our candidate for a non-BPS solution should have only seven nonvanishing Fourier amplitudes and no $D6$ brane charges (in the type IIA duality frame).

Let us now introduce the notation
\beq
y_0=e^{\phi_0},\qquad y_{j}=e^{\phi_{j}},\qquad \beta\equiv U-\frac{1}{2}(\phi_1+\phi_2+\phi_3), \qquad \alpha_j\equiv U+\frac{1}{2}\phi_j,
\label{phidefi}
\eeq
\noindent
with and $j=1,2,3$ (recall also that according to Eq.(\ref{ujkoord}) now $\phi\equiv\phi_0=2U$.) 
Now our equations take the form
\beq
\tilde{\chi}_{111}=\frac{1}{2}\dot{\beta},\qquad
\tilde{\chi}_{110}=\frac{i}{2}e^{-\phi_1}\dot{x}_1,\qquad
\tilde{\chi}_{101}=\frac{i}{2}e^{-\phi_2}\dot{x}_2,\qquad
\tilde{\chi}_{011}=\frac{i}{2}e^{-\phi_3}\dot{x}_3,
\label{gim1}
\eeq
\noindent
\beq
\tilde{\chi}_{001}=\frac{1}{2}(\dot{\alpha}_1-\dot{\alpha}_2-\dot{\alpha}_3),   \qquad
\tilde{\chi}_{010}=\frac{1}{2}(\dot{\alpha}_2-\dot{\alpha}_3-\dot{\alpha}_1),   \qquad
\tilde{\chi}_{100}=\frac{1}{2}(\dot{\alpha}_3-\dot{\alpha}_1-\dot{\alpha}_2).
\label{gim2}
\eeq
\noindent
Now using Eqs.(\ref{2x})-(\ref{8x})
with the further charge constraints $q_j=0$ , $q_0<0$, and $p^1,p^2,p^3>0$
one can see that the equations are precisely the ones found in the Appendix of the paper of Gimon et.al.\cite{seed} characterizing the seed solutions for the $D0-D4$ system.

We remark in closing that one can verify by an explicit calculation that all of the four algebraically independent four-qubit invariants $I_k, k=1,2,3,4$ are vanishing.
This means that the corresponding matrix $Q$ of conserved charges is nilpotent. Hence in the teminology of four-qubit entanglement we obtained the result: the relevant state $\vert\Lambda\rangle$, is a nilpotent one. However, unlike in the previous cases now neither of the qubits can be separated from the rest, hence $\vert\Lambda\rangle$ is also an entangled state. 

Notice however, that neither the order of nilpotency nor the particular entanglement type follows from our simple considerations.
It would be interesting to extend our analysis and identify the particular entanglement class to which $\vert\Lambda\rangle$ belongs case by case.  
It is important to realize in this respect,
that in our simplified considerations we have merely used {\it complex} four qubit states and the corresponding $SL(2,\mathbb C)^{\times 4}$ invariants.
However, we must recall that our state $\vert\Lambda\rangle$ also have to satisfy the reality condition of Eq. (\ref{realagain}).
The result of the implementation of this constraint is that in the black hole context we have to classify orbits under the group $SU(1,1)^{\times 4}$ which is merely a subgroup of the full group of admissible local operations\cite{Dur}.
Hence a full entanglement based understanding of black hole solutions in the STU model should rely on the classification of entanglement types of {\it real} four qubit states defined by Eq. (\ref{realagain}).
This classification should be founded on a study of $SU(1,1)^{\times 4}$ invariants. 
Clearly this reformulation would relate the known classification\cite{Bossard} of satic black hole solutions in the STU model in terms of nilpotent orbits to a similar one based on entanglement classes of the relevant {\it real} four qubit states.

\section{Conclusions}

In this paper we managed to understand the structure of extremal stationary spherically symmetric black hole solutions in the $STU$ model of $D=4$, $N=2$ supergravity
in terms of four-qubit systems.
Our analysis extended the results obtained in our previous papers based on three qubit systems\cite{Levay1,Levay3,Levszal}. 
The basic idea facilitating this $4$-qubit description was the fact 
that stationary solutions in $D=4$ supergravity
can be elegantly described by dimensional reduction along the time direction\cite{Gibbons}.
In this picture stationary solutions can be identified as solutions to a $D=3$ non-linear sigma model with target space being a symmetric space $G/H$ with $H$ {\it non-compact}.
The group $G$ extends the global symmetry group $G_4$ of $D=4$ supergravity, by also incorporating the Ehlers $SL(2,{\mathbb R})$.
In our specific case the $N=2$ STU model can be regarded as a consistent truncation of maximal $N=8$, $D=4$ supergravity with $G_4= E_{7(7)}$, truncating to $G_4= SL(2,{\mathbb R})^{\times 3}$.
Timelike reduction then yields the coset $E_{8(8)}/SO^{\ast}(16)$,
or in the case of the STU truncation the one ${\cal M}_3=SO(4,4)/SL(2,{\mathbb R})^{\times 4}$. We have shown that the four copies of $SL(2,{\mathbb R})$s  occurring in this coset can be reinterpreted as the group of local operations acting on four qubits subject to special reality constraints.
Here the fourth qubit which accounts for the Ehlers group played a special role.

The central object of our considerations was the {\it complex} $4$-qubit state $\vert\Lambda\rangle$ of Eqs. (\ref{ittalenyeg}) and (\ref{lambdastate}), also satisfying the reality condition Eq.(\ref{realagain}).
The amplitudes of this state of {\it odd parity} contain the right invariant one-forms  $e_{\alpha}$, $\alpha=0,1,2,3$ defined by Eqs.(\ref{eztitt})-(\ref{right}).
On the other hand the $8$ amplitudes of {\it even parity} 
are  just the $8$ amplitudes of the $3$-qubit state well-known from previous studies concerning the black hole qubit correspondence.
According to Eqs.(\ref{szerkezet}) and (\ref{nahalisten}) these amplitudes are related to well-known quantities of special geometry.
We have shown that the state $\vert\Lambda\rangle$ is connected to the line element on ${\cal M}_3$ via Eqs. (\ref{line}), (\ref{1v}) and (\ref{lineasinv}).
We also realized that this expression for the line element is minus the quadratic $4$-qubit $SL(2,\mathbb C)$ invariant  $I_1$ of Eq.(\ref{26}). 
After expressing the $8$ amplitudes of the embedded $3$-qubit state in terms of the conserved electric, magnetic and NUT charges as in Eqs.(\ref{ezittaklassz}), (\ref{naeztkapdki}) 
this invariant also has the physical interpretation as the BPS parameter\cite{Stelle}.
(For nonrotating solutions  this parameter is just the {\it extremality parameter}.)

We clarified the relationship between the warp factor, moduli and charge dependent $3$-qubit state of Eq.(\ref{3qubitallapot}), and (\ref{chi}) used in previous studies\cite{Levay1,Levay3,Levszal} and the $4$-qubit one $\vert\Lambda\rangle$.
Our considerations enabled a formal generalization for this state (see Eq.(\ref{naeztkapdki})) also valid for nonvanishing NUT charge.
Notice that for general stationary solutions the entanglement type of this state (i.e. the value of the three-tangle\cite{Kundu,Zelevinsky,Duff1}) is also depending on the Wilson lines ${\zeta}^I$ and ${\tilde{\zeta}}_I$.
This is in sharp contrast to the static case where the entanglement type is merely depending on the conserved electric and magnetic charges.

Note that one of the qubits of the state $\vert\Lambda\rangle$ was special.
We have seen that the special status of this qubit is related to the $R$-symmetry group arising from the resticted holonomy group of  the para quaternionic
K\"ahler space ${\cal M}_3$.
We realized that our special set of transformations, based on Hadamard and phase gates and permutations,
resulting in the explicit form for $\vert\Lambda\rangle$ correspond to the basis transformations similar to the ones rendering the quaternionic vierbein covariantly constant with respect to the spin connection\cite{Bossard}.

The separability properties of this special qubit are related to the solution being BPS or non-BPS.
We demonstrated within our formalism the observation of 
Bergshoeff et.al.\cite{Trigiante,Trigi} that static, extremal BPS and non-BPS-solutions with vanishing central charge\cite{Scherbakov} correspond to states for which one of the qubits is separable from the rest.
On the other hand using the non-BPS seed solution\cite{seed} for nonvanishing central charge we have shown that $\vert\Lambda\rangle$ in this case is entangled.
We revealed a connection between the classification of {\it nilpotent states} within the realm of quantum information theory and the similar classification of nilpotent orbits.
The details of this connection should be explored further.

It is amusing to see that nonextremal solutions should correspond to states
which are {\it semisimple}\cite{Djokovic}.
Since nilpotent states are rather exceptional among the $4$-qubit ones, semisimple states are the ones that represent genuine $4$-qubit entanglement.
According to our Appendix for such states at least one of the algebraically independent invariants (namely $I_1$ related to the extremality parameter) are non-vanishing.
Such states with special entanglement properties should correspond to nonextremal solutions of special kind.

Notice also in this respect that in the paper of Chemissany et.al.\cite{Trigi}
dealing with the full integration of black hole solutions in symmetric supergravity solutions the authors notice that in their Lax pair approach exactly non-extremal solutions are easier
to describe than extremal ones.
Such solutions correspond to {\it diagonalizable} initial conditions in terms of the Lax matrix.
On the other hand they note that the  initial conditions summarized in {\it nondiagonalizable} Lax matrices correpond to extremal BPS and non-BPS solutions,
however such Lax matrices represent a subset of measure zero within the space of Lax matrices. 
Within the context of the STU-model clearly semisimple states should correspond to diagonalizable Lax matrices, and nilpotent states to nondiagonalizable ones.
Hence there should be a correspondence between giving the Lax matrix at some initial time and specifying the entanglement properties of the corresponding entangled $4$-qubit state. 

Notice in particular the highly symmetrical nature of the genuine entangled $4$-qubit class (see Eq.(\ref{genuine}) of the Appendix and the structure of its invariants.)
This state is the $4$-qubit analogue of the famous GHZ state\cite{GHZ} familiar from $3$-qubit entanglement. Notice that for choosing $a,b,c,d\in{\mathbb R}$ this state automatically satisfies the reality condition of Eq.(\ref{realagain}). 
Acting on this state with $SU(1,1)^{\times 4}$ transformations preserving this reality condition and also the values af the algebraically independent invariants results in a state $\vert\Lambda\rangle$ containing $16$ real parameters. 
Using this parametrization and the black hole qubit correspondence it would be amusing to find a corresponding highly symmetrical non-extremal solution.

Recall also the classification of black hole solutions in the STU model in terms of three qubit entanglement classes as given by Kallosh and Linde\cite{Linde}.
In this paper the authors noticed a similarity between the classification of {\it complex} three qubit states\cite{Dur} and the corresponding classification of small and large black holes in the STU model related to {\it real} three qubit ones.
In the light of our results we might substantially generalize this interesting result.
Indeed, by embedding the usual three-qubit picture into the four qubit one as described here, we also have the possibility to include such notions as BPS and non-BPS , extremal and non extremal solutions into an entanglement based picture.
As we have seen the extremality parameter is related to the quadratic four qubit invariant. Extremal black holes are characterized by the vanishing of this quantity.
Though the remaining four qubit invariants are all vanishing for the known extremal solutions, but such solutions are still distinguished by their entanglement properties. For BPS and non-BPS solutions with vanishing central charge one of the qubits is separable from the rest, and for the $Z\neq 0$ case none of the qubits is separable.
Since the states describing such solutions are {\it real} (i.e. they are satisfying the reality condition of Eq.(\ref{realagain})) in order to classify their orbit structure we also have to include some additional $SU(1,1)^{\times 4}$ invariants. 
We know that the classes in question for extremal solutions are just the nilpoten orbits classified in the paper of Bossard et. al.\cite{Bossard}, hence these classes might be distinguished by additional $SU(1,1)^{\times 4}$ invariants whose physical meaning is still to be clarified.
On the other hand the three qubit part of our four qubit states is classified by the value of Cayleys hyperdeterminant i.e. the three-tangle\cite{Kundu}. 
For small black holes this invariant is vanishing and for large ones it is nonzero and its value is proportional to the black hole entropy.

Finally notice that we have deliberately emphasized the possibility 
for reformulating the well-known results of the STU model in a suggestive permutation invariant language (see e.g. Eqs. (\ref{line1}-(\ref{norma})). Though instructive, this language is deceptive due to the special role we have attached to our first qubit via the use of the Ehlers $SL(2,\mathbb R)$. However, since the quadratic invariant of Eq.(\ref{lineasinv}) related to the line element and the extremality parameter is a permutation invariant quantity one might speculate whether there is a further possibility for embedding the STU model into an even greater picture where permutation symmetry is manifest. 
In this respect an exciting possibility is to find the physical relevance (if any) of the permutation invariant quantity of Eq.(\ref{hyper4}) i.e. the four qubit generalization of Cayley's hyperdeterminant (the "four-tangle").

\section{Appendix}

A four qubit state can be written in the form

\beq
\vert {\Psi}\rangle =\sum_{i_3i_2i_1i_0=0,1}{\Psi}_{i_3i_2i_1i_0}\vert i_3i_2i_1i_0\rangle,\quad \vert
i_3i_2i_1i_0\rangle\equiv \vert i_3\rangle\otimes\vert i_2\rangle\otimes\vert i_1\rangle\otimes
\vert i_0\rangle\in V_3\otimes V_2\otimes V_1
\otimes V_0,
\eeq
\noindent
where $V_{3,2,1,0}\equiv {\mathbb C}^2$.
Let the subgroup of stochastic local operations and classical communication\cite{Dur}
representing admissible fourpartite protocols be $SL(2, {\mathbb C})^{\times 4}$
acting on $\vert\Psi\rangle$ as

\beq
\vert\Psi\rangle\mapsto (S_3\otimes S_2\otimes S_1\otimes S_0)                  \vert\Psi\rangle,   \quad                                             S_{\alpha}\in SL(2, {\mathbb C} ) ,\quad\alpha=0,1,2,3.                                   \eeq                                                                            \noindent                                                                                                                                                       Our aim in this appendix is to give a unified description of four-qubit states taken together with their SLOCC transformations and their associated invariants. As we will see states and transformations taken                               together                                                                         can be described in a unified manner using the group $SO(4,4,{\mathbb C})$.    

Let us introduce the $2\times 2$ matrices
                                                                                \beq
										E_{00}=\begin {pmatrix}1&0\\0&0\end {pmatrix},\quad
										E_{01}=\begin {pmatrix}0&1\\0&0\end {pmatrix},\quad
										E_{10}=\begin {pmatrix}0&0\\1&0\end {pmatrix},\quad
										E_{11}=\begin {pmatrix}0&0\\0&1\end {pmatrix}.
										\eeq
										\noindent
										Then we arrange the $16$ complex amplitudes appearing in ${\Psi}_{i_3i_2i_1i_0}$
										in a $4\times 4 $ matrix in three different ways

										\beq
										D_1(\Psi)=\sum_{i_3i_2i_1i_0=0,1}{\Psi}_{i_3i_2i_1i_0}E_{i_3i_1}\otimes E_{i_2i_0}
										\equiv\begin {pmatrix}{\Psi}_{0000}&{\Psi}_{0001}&{\Psi}_{0010}&{\Psi}_{0011}\\
										{\Psi}_{0100}&{\Psi}_{0101}&{\Psi}_{0110}&{\Psi}_{0111}\\
										{\Psi}_{1000}&{\Psi}_{1001}&{\Psi}_{1010}&{\Psi}_{1011}\\
										{\Psi}_{1100}&{\Psi}_{1101}&{\Psi}_{1110}&{\Psi}_{1111}\end {pmatrix},
\label{elsoelrendez}										\eeq
										\noindent

\beq
D_2(\Psi)=\sum_{i_3i_2i_1i_0=0,1}{\Psi}_{i_3i_2i_1i_0}E_{i_3i_2}\otimes E_{i_1i_0}=
\begin {pmatrix}{\Psi}_{0000}&{\Psi}_{0001}&{\Psi}_{0100}&
{\Psi}_{0101}\\
{\Psi}_{0010}&{\Psi}_{0011}&{\Psi}_{0110}&{\Psi}_{0111}\\
{\Psi}_{1000}&{\Psi}_{1001}&{\Psi}_{1100}&{\Psi}_{1101}\\
{\Psi}_{1010}&{\Psi}_{1011}&{\Psi}_{1110}&{\Psi}_{1111}\end {pmatrix}=          \begin{pmatrix}
X&Y\\W&Z\end{pmatrix},
\eeq
\noindent

\beq
D_3(\Psi)=\sum_{i_3,i_2,i_1,i_0=0,1}{\Psi}_{i_3i_2i_1i_0}E_{i_2i_3}\otimes E_{i_1i_0}=
\begin{pmatrix}X&W\\Y&Z\end{pmatrix}
\label{d3}
\eeq
\noindent
where the $2\times 2$ matrices $X,Y,W,Z$ are introduced merely to illustrate
the block structure of the relevant matrices.
The first matrix is obtained by arranging the components of
$X,Y,W,Z$ as the first, second, third and fourth rows.
Notice that the arrangement $D_1$ of Eq.(\ref{elsoelrendez}) is our one of Eq.(\ref{ime}). 
Clearly changing $D_1$ to $D_2$
or to $D_3$ corresponds to the two generators of the permutation group $S_3$ acting on the last three qubits. 
Applying such permutations to the qubits corresponds to a similar permutation of the entries $s_3,s_2,s_1$ of Eq.(\ref{fontos}) resulting in the matrices ${\cal D}(s_3,s_1,s_2,s_0;D_2)$, and 
${\cal D}(s_2,s_3,s_1,s_0;D_3)$. 
These permutations give rise to alternative forms for the matrix exponentials of Eq.(\ref{1f}-\ref{cayleys}) with special roles attached to the second and the third qubit respectively.

Our matrix ${\cal D}\equiv{\cal D}(s_3,s_2,s_1,s_0; D_1)$ of Eq.(\ref{fontos})
in the parametrization used in Bossard et.al.\cite{Bossard} takes the following form

\beq
\begin{pmatrix}H_3+H_2&E_2&E_3&0&-F_{q_1}&-E_{p^1}&-F_{q_0}&-E_{p^0}\\
F_2&H_3-H_2&0&E_3&F_{p^3}&-E_{q_3}&F_{q_2}&E_{p^2}\\
F_3&0&H_2-H_3&E_2&F_{p^2}&-E_{q_2}&F_{q_3}&E_{p^3}\\
0&F_3&F_2&-H_3-H_2&F_{p^0}&-E_{q_0}&-F_{p^1}&E_{q_1}\\
-E_{q_1}&E_{p^3}&E_{p^2}&E_{p^0}&H_1+H_0&E_0&E_1&0\\
-F_{p_1}&-F_{q_3}&-F_{q_2}&-F_{q_0}&F_0&H_1-H_0&0&E_1\\
-E_{q_0}&E_{q_2}&E_{q_3}&-E_{p^1}&F_1&0&H_0-H_1&E_0\\
-F_{p^0}&F_{p^2}&F_{p^3}&F_{q_1}&0&F_1&F_0&-H_1-H_0\end{pmatrix}.
\label{explicitbossard}
\eeq
\noindent
It can be checked that the matrix $S{\cal D}S^T$
where
\beq
S\equiv\begin{pmatrix}1&0&0&0&0&0&0&0\\
0&0&-1&0&0&0&0&0\\
0&0&0&0&1&0&0&0\\
0&0&0&0&0&0&-1&0\\
0&0&0&-1&0&0&0&0\\
0&-1&0&0&0&0&0&0\\
0&0&0&0&0&0&0&-1\\
0&0&0&0&0&-1&0&0\end{pmatrix}
\label{smatrix}
\eeq
\noindent
is just the one used in Eq.(4.6) of that paper.
This matrix also relates our matrix $G$ of Eq.(\ref{GG}) to the usual $SO(4,4)$
invariant one
\beq
\eta=\begin{pmatrix}{\bf 0}&{\bf 1}\\{\bf 1}&{\bf 0}\end{pmatrix},\qquad
{\bf 1}=I\otimes I.
\eeq
\noindent
Relating the upper right block of Eq.(\ref{explicitbossard}) to the $4\times4$ matrix $Dg$ of Eq.(\ref{fontos}) shows that in this parametrization

\beq
D=D_1=\begin{pmatrix}-E_{p^0}&F_{q_0}&E_{p^1}&-F_{q_1}\\
                   E_{p^2}&-F_{q_2}&E_{q_3}&F_{p^3}\\
		   E_{p^3}&-F_{q_3}&E_{q_2}&F_{p^2}\\
		   E_{q_1}&F_{p^1}&E_{q_0}&F_{p^0}\end{pmatrix}
\eeq
\noindent
which justifies our parametrization of $\zeta^IE_{q_I}+\tilde{\zeta}_IE_{p^I}$
used in Eq.(\ref{vissza}).
		   
Let us discuss now the structure of four-qubit $SL(2,\mathbb C)^{\times 4}$ invariants\cite{Luque,Levay41,Djokovic,Osterloh}.
The number of algebraically independent four-qubit invariants is four.
We have one quadratic, two quartic, and one sextic invariant.
In our recent paper\cite{Levay41} we investigated the structure of these invariants in the special frame where two of our qubits played a distinguished role.
Clearly this is the case in the black hole context, since one of the special qubits is associated with the Ehlers-group and the choice of the other is just a 
matter of convention related to the special 
choice $D_1$, $D_2$ or $D_3$ of Eqs.(\ref{elsoelrendez}-\ref{d3}).

To an arbitrary state
\beq
\vert\Lambda\rangle =\sum_{i_3i_2i_1i_0=0,1}\Lambda_{i_3i_2i_1i_0}\vert i_3i_2i_1i_0\rangle,
\eeq
\noindent
we can associate the $4\times 4$ matrix
\beq
\Lambda\equiv
\begin {pmatrix}{\Lambda}_{0000}&{\Lambda}_{0001}&{\Lambda}_{0010}&{\Lambda}_{0011}\\
{\Lambda}_{0100}&{\Lambda}_{0101}&{\Lambda}_{0110}&{\Lambda}_{0111}\\
{\Lambda}_{1000}&{\Lambda}_{1001}&{\Lambda}_{1010}&{\Lambda}_{1011}\\
{\Lambda}_{1100}&{\Lambda}_{1101}&{\Lambda}_{1110}&{\Lambda}_{1111}\end{pmatrix}
\equiv
\begin{pmatrix}A^1&B^1&C^1&D^1\\
A^2&B^2&C^2&D^2\\
A^3&B^3&C^3&D^3\\
A^4&B^4&C^4&D^4\end{pmatrix},
\label{fourvectors}
\eeq
\noindent
or four four-vectors.
The splitting of the amplitudes of $\vert\Lambda\rangle$ into four four-vectors reflects our special choice for the distinguished qubits compatible with our conventions.
Now we introduce on the vector space ${\mathbb C}^4\simeq {\mathbb C}^2\times {\mathbb C}^2$
corresponding to the third and fourth qubit a symmetric bilinear form ${\bf g}:{\mathbb C}^4\times{\mathbb C}^4\to{\mathbb C}$ with matrix representation: $g\equiv{\varepsilon}\otimes {\varepsilon}$. This means that we have
an $SL(2,\mathbb C)^{\times 2}$ invariant quantity with the explicit form
\beq
g(A,B)\equiv g_{\alpha\beta}A^{\alpha}B^{\beta}=A_{\alpha}B^{\alpha}=A\cdot B= A^1B^4-A^2B^3-A^3B^2+A^4B^1.
\label{explicitalak}
\eeq
\noindent

We can also introduce a {\it dual four-qubit state}
\beq
\vert\lambda\rangle=\sum_{i_3i_2i_1i_0=0,1}\lambda_{i_3i_2i_1i_0}\vert i_3i_2i_1i_0\rangle
\eeq
\noindent
with the associated matrix
\beq
\lambda\equiv
\begin {pmatrix}{\lambda}_{0000}&{\lambda}_{0001}&{\lambda}_{0010}&{\lambda}_{0011}\\                                                                           {\lambda}_{0100}&{\lambda}_{0101}&{\lambda}_{0110}&{\lambda}_{0111}\\           {\lambda}_{1000}&{\lambda}_{1001}&{\lambda}_{1010}&{\lambda}_{1011}\\           {\lambda}_{1100}&{\lambda}_{1101}&{\lambda}_{1110}&{\lambda}_{1111}\end{pmatrix}\equiv                                                                          \begin{pmatrix}a^1&b^1&c^1&d^1\\                                                a^2&b^2&c^2&d^2\\                                                               a^3&b^3&c^3&d^3\\                                                               a^4&b^4&c^4&d^4\end{pmatrix},                                                   \label{fourvectorsdual}                                                         \eeq                                                                            \noindent
where
\beq
a^{\alpha}={\epsilon}^{\alpha\beta\gamma\delta}B_{\beta}C_{\gamma}D_{\delta},\quad
b^{\beta}={\epsilon}^{\alpha\beta\gamma\delta}A_{\alpha}C_{\gamma}D_{\delta}\quad
c^{\gamma}={\epsilon}^{\alpha\beta\gamma\delta}A_{\alpha}B_{\beta}D_{\delta}\quad
d^{\delta}={\epsilon}^{\alpha\beta\gamma\delta}A_{\alpha}B_{\beta}C_{\gamma}.
\label{dual}
\eeq
\noindent
Here  ${\epsilon}^{1234}=+1$, and indices are lowered by the matrix of $g$.
Notice that the amplitudes of the dual four-qubit state are {\it cubic}
in the original ones.
Such dual states were first introduced in Ref.{\cite{Levay41}}, and were later
defined differently in the three-qubit context by Borsten et.al.\cite{Borsten2}. These dual states have also made their debut to the physics of black holes admitting Freudenthal or Jordan duals\cite{Borsten2,Borsten3}.

Using these definitions we define the quadratic and sextic invariants as
\beq
I_1\equiv \frac{1}{2}(A\cdot D-B\cdot C),\qquad I_3\equiv \frac{1}{2}(a\cdot d-b\cdot c).
\label{26}
\eeq
\noindent
(The labelling convention and normalization for our invariants will be clarified below.)
This form of the sextic invariant is deceptively simple. Its explicit form in terms of the dot product of Eq.(\ref{explicitalak}) is
\beq
2I_3={\rm Det}\begin{pmatrix}A\cdot A&A\cdot B&A\cdot D\\
A\cdot C&B\cdot C&C\cdot D\\
A\cdot D&B\cdot D&D\cdot D\end{pmatrix}-
{\rm Det}\begin{pmatrix}A\cdot B&B\cdot B&B\cdot C\\
A\cdot C&B\cdot C&C\cdot C\\
A\cdot D&B\cdot D&C\cdot D\end{pmatrix}.
\label{sexkiirva}
\eeq
\noindent
We also recall that the explicit form of $I_1$ is hiding its permutation invariance. A permutation invariant form is the one we used in Eq.(\ref{lineasinv}),
i.e. we have $ds^2=-I_1(\vert\Lambda\rangle)$ where $\vert\Lambda\rangle$ is the special state of Eq.(\ref{ittalenyeg}).
Moreover, though the expression of $I_3$ of Eq.(\ref{26}) is similar to the one of $I_1$ the invariant $I_3$ is {\it not} invariant under the permutation of the qubits.

Now we turn to the structure of quartic invariants. We have two of them and the simplest is the obvious expression
\beq
I_4\equiv {\rm Det}\Lambda
\label{detinv}
\eeq
\noindent
i.e. the determinant of the $4\times 4$ matrix of Eq.(\ref{fourvectors}).
In order to present the definition of our last invariant we define
separable bivectors of the form
\beq
\Pi_{\mu\nu\alpha\beta}\equiv \Lambda_{\mu\alpha}\Lambda_{\nu\beta}-\Lambda_{\mu\beta}\Lambda_{\nu\alpha},\qquad \alpha,\beta,\mu,\nu=1,2,3,4.
\label{bivectors}
\eeq
\noindent
Here our labelling convention $\Lambda_{\mu\alpha}$ indicates that $\alpha=1,2,3,4$ identifies the four-vector in question (i.e. $A,B,C$ or $D$ of Eq.(\ref{fourvectors})), and the label $\mu=1,2,3,4$ refers to the component of the particular vector. 
Now our last invariant is the quartic combination
\beq
I_2=\frac{1}{6}\Pi_{\mu\nu\alpha\beta}\Pi^{\mu\nu\alpha\beta}.
\label{last4}
\eeq
\noindent
For the explicit form of this invariant we introduce the $\cdot$ product of two separable bivectors as
\beq
(A\wedge B)\cdot(C\wedge D)\equiv 2(A\cdot C)(B\cdot D)-2(A\cdot D)(B\cdot C).
\label{defike}
\eeq
\noindent
Then the explicit form is
\beq
I_2=\frac{1}{6}\left[(A\wedge B)\cdot(C\wedge D)+(A\wedge C)\cdot(B\wedge D)-\frac{1}{2}(A\wedge D)^2
-\frac{1}{2}(B\wedge C)^2\right].
\label{explutolso}
\eeq
\noindent

Let us now present the reason for considering these particular combinations
for the set of algebraically independent $SL(2,\mathbb C)^{\times 4}$ invariants.
Let us consider the matrix
\beq
\Omega\equiv {\Lambda}^Tg\Lambda g.
\label{Omega}
\eeq
\noindent
Then the characteristic polynomial of this $4\times 4$ matrix 
is
\beq
{\rm Det}({\bf 1}t-\Omega)=t^4-4I_1t^3+6I_2t^2-4I_3t+I_4^2.
\label{poli}
\eeq
\noindent
Clearly we have
\beq
I_1=\frac{1}{4}{\rm Tr}\Omega,\qquad I_2=\frac{1}{12}[({\rm Tr}\Omega)^2-{\rm Tr}\Omega^2],
\eeq
\noindent
\beq
I_3=\frac{1}{24}[({\rm Tr}\Omega)^3-3{\rm Tr}\Omega{\rm Tr}\Omega^2+2{\rm Tr}\Omega^3],\qquad (I_4)^2={\rm Det}\Omega.
\eeq
\noindent
This form of writing our invariants is related to the fact that there is a $1-1$ correspondence between the $SL(2,{\mathbb C})^{\times 4}$ orbits of four-qubit states and the $SO(4,\mathbb C)\times SO(4,\mathbb C)$ ones of $4\times 4$ matrices.

The polynomial of Eq.(\ref{poli}) first appeared in Ref.\cite{Levay41}  its role as a characteristic polynomial has been emphasized in Ref.\cite{Djokovic}.
The discriminant of this fourth order polynomial is the hyperdeterminant\cite{Zelevinsky} $D_4$ of the
$2\times 2\times 2\times 2$ hypercube $\Lambda_{i_3i_2i_1i_0}$.
It is a polynomial of degree $24$ in the $16$ amplitudes and has 2894276 terms\cite{Debbie}. It can be shown\cite{Luque,Levay41} that $D_4$ can be expressed in terms of our fundamental invariants in the form 
\beq                                                                            256D_4=S^3-27T^2                                                                
\label{hyper4}
\eeq
\noindent
where                                                                           \beq
S=(I_4^2-I_2^2)+4(I_2^2-I_1I_3),\quad T=(I_4^2-I_2^2)(I_1^2-I_2)+(I_3-I_1I_2)^2.
\eeq
\noindent

In closing this appendix we briefly discuss some results on  the full  classification of entanglement classes for four qubits\cite{Verstraete,Djokovic}.
By entanglement classes we mean orbits under $SL(2,{\mathbb C})^{\times 4}\cdot {\rm Sym}_4$ where ${\rm Sym}_4$ is the symmetric group on four symbols.
The basic result states that four qubits can be entangled in
nine different ways\cite{Verstraete,Djokovic}. It is to be contrasted with the two entanglement classes\cite{Dur}
obtained for three qubits.

Let us consider the matrix
\beq
{\cal R}_{\Lambda}\equiv\begin{pmatrix}0&\Lambda g\\-\Lambda^Tg&0\end{pmatrix}.
\label{rmatrix}
\eeq
\noindent
If $\Lambda$ is the special matrix of Eq.(\ref{ittalenyeg}) used in the black hole context ${\cal R}_{\Lambda}$ is just $2{\cal P}_{\ast}^{\prime}$ of Eq.(\ref{1v}).
If the matrix ${\cal R}_{\Lambda}$ is diagonalizable under the action  
\beq
{\cal R}_{\Lambda}\mapsto S{\cal R}_{\Lambda}S^{-1},\qquad S=\begin{pmatrix}S_3\otimes S_2&0\\0&S_1\otimes S_0\end{pmatrix},\qquad S_{\alpha}\in SL(2,{\mathbb C})
\eeq
\noindent 
we say that the corresponding four-qubit state $\vert\Lambda\rangle$ is {\it semisimple}. 
If ${\cal R}_{\Lambda}$ is {\it nilpotent} then we call the corresponding state $\vert\Lambda\rangle$ nilpotent too.
It is known that a nilpotent orbit is {\it conical} i.e. if $\vert\Lambda\rangle$ is an element of the orbit then $t\vert\Lambda\rangle$ is also an element for all nonzero complex numbers $t$. Hence a  nilpotent 
orbit is also a $GL(2,{\mathbb C})^{\otimes 4}$ orbit.
(Recall that our ${\cal P}^{\prime}_{\ast}$ is in the $GL(2,{\mathbb C})^{\otimes 4}$ orbit of the original ${\cal P}$ of Eq.(\ref{explicit}.)
It is clear that for nilpotent states all of our algebraically independent invariants are zero. 
These are the states we associated to extremal black hole solutions of BPS and non-BPS type.

A generic semisimple state of four qubits can always be transformed
to the form\cite{Verstraete}
\begin{eqnarray}
\vert G_{abcd}\rangle&=&\frac{a+d}{2}(\vert 0000\rangle +\vert 1111\rangle)
+\frac{a-d}{2}(\vert 0011\rangle +\vert 1100\rangle)\nonumber\\&+&
\frac{b+c}{2}(\vert 0101\rangle +\vert 1010\rangle)+
\frac{b-c}{2}(\vert 0110\rangle +\vert 1001\rangle),
\label{genuine}
\end{eqnarray}
\noindent
where $a,b,c,d$ are complex numbers. This class corresponds to the so called GHZ class
found in the three-qubit case\cite{Dur}.
For this state the reduced density matrices obtained by tracing out all but one
of the qubits are proportional to the identity. This is the state with maximal four-partite
entanglement.

Another interesting property of this state is that it does not contain true three-partite entanglement.
A straightforward calculation shows that the values of our invariants $(I_1,I_2,I_3,I_4)$ occurring for the state $\vert G_{abcd}\rangle$ representing the generic class are

\beq
I_1=\frac{1}{4}[a^2+b^2+c^2+d^2],\quad
I_2=\frac{1}{6}[(ab)^2+(ac)^2+(ad)^2+(bc)^2+(bd)^2+(cd)^2],
\eeq
\noindent
\beq
I_4=\frac{1}{4}[(abc)^2+(abd)^2+(acd)^2+(bcd)^2],\quad I_3=abcd,
\eeq
\noindent
hence the values of the invariants $(4I_1,6I_2,4I_3,I_4^2)$ are given in terms of the elementary symmetric polynomials in the
variables $(x_1,x_2,x_3,x_4)=(a^2,b^2,c^2,d^2)$.
On the generic class $\vert G_{abcd}\rangle$ the value of $D_4$ can be expressed as\cite{Luque,Levay41}
\beq
D_4=\frac{1}{256}\Pi_{i<j}(x_i-x_j)^2=\frac{1}{256}V(a^2,b^2,c^2,d^2)^2,
\eeq
\noindent
where $(x_1,x_2,x_3,x_4)\equiv (a^2,b^2,c^2,d^2)$ and $V$ is the Vandermonde determinant.

\section{Acknowledgement}
The author would like to express his gratitude to the warm hospitality at the Center for Interdisclipinary Research (ZiF), University of Bielefeld, Germany, giving home to the Cooperation Group "Finite Projective Geometries", where the basic ideas of this work have been conceived.


\begin{thebibliography}{}

\bibitem{Duff2} M. J. Duff, Phys. Rev. D{\bf 76} 025017 (2007), arXiv:hep-th/0601134.
\bibitem{Linde} R. Kallosh, A. Linde, Phys. Rev. D{\bf 73} 104033 (2006), arXiv:
0602061.
\bibitem{Levay1} P. L\'evay, Phys. Rev. D{\bf 74}, 024030 (2006), arXiv:0603136.
\bibitem{DF1} M. J. Duff, S. Ferrara, Phys. Rev. D{\bf 76} 025018 (2007), arXiv:quant-ph0609227.
\bibitem{Levay2} P. L\'evay, Phys. Rev. D{\bf 75} 024024 (2007), arXiv:hep-th/0610314.
\bibitem{DF2} M. J. Duff, S. Ferrara, Phys. Rev. D{76} 124023 (2007), arXiv:0704.0507[hep-th].
\bibitem{Levay3} P. L\'evay, Phys. Rev. {\bf D76}, 106011 (2007), arXiv:0708.2799 [hep-th].
\bibitem{Scherbakov} S. Bellucci, A. Marrani, E. Orazi, A. Scherbakov, Phys. Lett. B{655} 185 (2007), 
arXiv:0707.2730[hep-th].
\bibitem{Borsten1} L. Borsten, D. Dahanayake, M. J. Duff, W. Rubens, H. Ebrahim,Phys. Rev. Lett. {\bf 100} 251602 (2008), arXiv:0802.0840[hep-th],
\bibitem{Borsten2}L. Borsten, Fortschr. Phys. {\bf 56} (7-9) 842 (2008),
L. Borsten, D.Dahanayake, M. J. Duff, H. Ebrahim, W. Rubens,
Phys.Rev.A{\bf 80} 032326 (2009),
arXiv:0812.3322[quant-ph],
L. Borsten, D. Dahanayake, M. J. Duff, W. Rubens,
Phys. Rev. D{\bf 80} 026003 (2009), arXiv:0903.5517[hep-th].
\bibitem{stu} S. Bellucci, S. Ferrara, A. Marrani, A. Yerayan,
Entropy 2008 Vol. 10(4), p. 507-555,
arXiv:0807.3503[hep-th].
\bibitem{Borsten3} L. Borsten, D. Dahanayake, M. J. Duff, H. Ebrahim and W. Rubens, Physics Reports, {\bf 471} 113 (2009), arXiv:0809.4685[hep-th].
\bibitem{LVS} P. L\'evay , M. Saniga and P. Vrana, Phys. Rev. D{\bf 78}, 124022 (2008), arxiv:0808.3849[quant-ph], P. L\'evay, M. Saniga, P. Vrana, P. Pracna,
Phys. Rev. D{\bf 79} 084036, (2009), arXiv:0903.0541[hep-th],
P. L\'evay and P. Vrana, Phys. Rev. A{\bf 78} 022329 (2008), arXiv:0806.4076[quant-ph], P. Vrana and P. L\'evay, 
J. Phys. A: Math. Theor. {\bf 42} 285303 (2009), arXiv:0902.2269[quant-ph]. 
\bibitem{Duff1} M. J. Duff, J. T. Liu and J. Rahmfeld, Nucl. Phys. {\bf B459},125 (1996), arXiv:hep-th/9508094,
K. Behrndt, R. Kallosh, J. Rahmfeld, M. Shmakova and W. K. Wong, Phys. Rev. D {\bf 54}, 6293 (1996).arXiv:hep-th/9608059,
\bibitem{attractor} S. Ferrara, R. Kallosh and A. Strominger, Phys. Rev. D{\bf 52} 5412 (1995), S. Ferrara nad R. Kallosh, Phys. Rev. D{\bf 54} 1514 (1996), R. Kallosh, Phys. Rev. D{\bf 54}, 1525 (1996),A. Strominger, Phys. Lett. {\bf B383} 39 (1996), S. Ferrara, G. W. Gibbons and R. Kallosh, Nucl. Phys. {\bf B500} 75 (1997), arXiv:hep-th/9702103.
\bibitem{Levszal} P. L\'evay and Sz. Szalay, "The attractor mechanism as a distillation procedure", arXiv:1004.2346[hep-th]. 
\bibitem{Kundu} V. Coffman, J. Kundu, and W. K. Wootters, Phys. Rev. A{\bf 61},
052306 (2000).
\bibitem{Cayley} A. Cayley, Camb. Math. J. {\bf 4}, 193 (1845).
\bibitem{Zelevinsky} I. M. Gel'fand, M. M. Kapranov and A. V. Zelevinsky,
{\it Discriminants, resultants and multidimensional determinants}, Birkh\"auser,
 Boston 1994.
\bibitem{GHZ} D. M. Greenberger, M. Horne. A. Zeilinger, {\it Bell's theorem.}, er. M. Kafatos, Kluwer, Dordrecht 69 (1989).
\bibitem{graph} M. Hein, W. D\"ur, J. Eisert, R. Raussendorf, M. Van den Nest, and H. J. Briegel, arXiv:quant-ph/0602096.
\bibitem{seed}K. Hotta and T. Kubota, Prog. Theor. Phys. {\bf 118N5}, 969 (2007),arXiv:0707.4554[hep-th],
E. G. Gimon, F. Larsen and J. Sim\'on, J. High Energy
Physics 01, 040 (2008), arXiv:0710.4967[hep-th],
Rong-Gen Cai and Da-Wei Pang,
J. High Energy Physics 01, 046 (2008),
arXiv:0712.0217[hep-th].
\bibitem{flat} S. Nampuri, P. K. Tripathy and S. Trivedi, Journal of High Energy Physics {\bf 0708} 054 (2007), arXiv:0705.4554[hep-th], S. Ferrara and A. Marrani, Phys. Lett. {\bf B652} 111 (2007), arXiv:0706.1667[hep-th].
\bibitem{fake}
D. Z. Freedman, C. Nunez, M. Schnabl and K. Skenderis, Phys. Rev. D{\bf 69} 104027 (2004), arXiv:hep-th/0312055, A. Celi, A. Ceresole, G. Dall'Agata, A. Van Proeyen and M. Zagerman, Phys. Rev. D{\bf71} 045009 (2005),
A. Ceresole and G. Dall'Agata, Journal of High Energy Physics {\bf 0703}, 110 (2007), arXiv:hep-th/0702088,
L. Andrianopoli, R. D'Auria, E. Orazi, M. Trigiante, Journal of High energy Physics {\bf 0711} 032 (2007), arXiv:0706.0712[hep-th].
\bibitem{NUT} E. T. Newman, L. Tamburino, and T. Unti, J. Math. Phys. {\bf 4} 915 (1963).
\bibitem{Ehlers} J. Ehlers, "Konstruktionen und Characterisierung von L\"osungen der Eimsteinischen Gravitationsfeldgleichungen", PhD Thesis, Hamburg University (1957). R. Geroch, J. Math. Phys. {\bf 12} 918 (1971).
\bibitem{Bossard} G. Bossard, Y. Michel and B. Pioline,
Journal of High Energy Physics 1001:038 (2010), arXiv:0908.1742.
\bibitem{Stelle}G. Bossard, H. Nicolai and K.S. Stelle, Journal of High energy Physics, 0907:003 (2009), arXiv:09024438.
\bibitem{Stelle2} G. Bossard, H. Nicolai and K. S. Stelle,
Gen. Rel. Grav. {\bf 41}, 1367 (2009), arXiv:0809.5218.
\bibitem{pioline2} M. G\"unaydin, A. Neitzke, B. Pioline and A. Waldron, Journal of High Energy Physics {\bf 09}, 056 (2007).
\bibitem{Trigiante} E. Bergshoeff, W. Chemissany, A. Ploegh, M. Trigiante and T.Van Riet, Nucl. Phys. {\bf B812} 343 (2009), arXiv:0806.2310,
\bibitem{Trigi}
W. Chemmisany, J. Roseel, M. Trigiante and T. Van Riet, arXiv:0903.2777.
\bibitem{Padi} D. Gaiotto, W. W. Li and M. Padi, Journal of High Energy
 Physics
\bibitem{Galtsov} G. Clement and D. V. Galtsov, Phys. Rev. D{\bf 54}, 6136 (1996), arXiv:hep-th/9607043.
{\bf 12} 093 (2007), arXiv:0710.1638 [hep-th].
\bibitem{Gibbons} P. Breitenlohner, D. Maison, and G. W. Gibbons, Commun. Math. Phys. {\bf 120} 295 (1988), P. Breitenlohner, D. Maison, Commun. Math. Phys. {\bf 209}, 785 (2000).
\bibitem{Pioline1} M. G\"unaydin, A. Neitzke, B. Pioline and A. Waldron, Journal
 of High Energy Physics {\bf 0709} 056 (2007).
\bibitem{Luque} J-G Luque and J-Y Thibon, Phys. Rev. A{\bf 67} 042303 (2003)
\bibitem{Dauria} A. Strominger, Commun. Math. Phys. {\bf 133} 163 (1990).
A. Ceresole, R. D'Auria and S. Ferrara, Nucl. Phys. Suppl. {\bf 46} 67 (1996).
\bibitem{Bates} B. Bates and F. Denef, arXiv: hep-th/0304094.
\bibitem{Feri} S. Ferrara and S. Sabharwal, Nucl.Phys. {\bf B332} 317, (1990).
\bibitem{Levay41} P. L\'evay, Journal of Physics A{\bf 39} 9533 (2006).
\bibitem{Dur} W. Dur, G. Vidal, and J. I. Cirac, Phys. Rev. A{\bf 62}, 062314 (2
000).
\bibitem{Beng} I. Bengtsson and K. Zyczkowski, {\it Geometry of quantum states},
   Cambridge (2006).
\bibitem{Djokovic} O. Chterental and D. Z. Dokovic, {\it Linear Algebra Research Advances} (Nova Science, Hauppauge, NY, 2007), Chap. 4. p. 133. arXiv:quant-ph/0612184.
\bibitem{Osterloh} D. Z. Dokovic and A. Osterloh, J. Math. Phys. {\bf 50}, 033509 (2009).
\bibitem{Debbie} P. Huggins, B. Sturmfels, J. Yu and D. S. Yuster, Math. Comp. {\bf 77} 1653 (2008).
\bibitem{Verstraete} F. Verstraete, J. Dehaene, B. De Moor and H. Verschelde, Phys. Rev. A{\bf 65} 052112 (2002).
\end{thebibliography}
\end{document}